\documentclass[9pt,preprint2]{aastex}

\shorttitle{Chemical Abundances in Symbiotic Stars}
\shortauthors{L\"{u} et al.}

\begin{document}


\title{Chemical Abundances in Symbiotic Stars }
\author{Guoliang L\"{u}\altaffilmark{1}$^\dagger$, Chunhua Zhu\altaffilmark{1,3},
Zhanwen Han\altaffilmark{2}, Zhaojun Wang\altaffilmark{1,3}}
\email{$^\dagger$GuoliangLv@gmail.com}
\altaffiltext{1}{Department of Physics, Xinjiang University, Urumqi,
830046, China.} \altaffiltext{2}{National Astronomical Observatories
/ Yunnan Observatory, the Chinese Academy of Sciences, P.O.Box 110,
Kunming, 650011, China}\altaffiltext{3}{School of Science, Xi'an
Jiaotong University, Xi'an, 710049, China}





\begin{abstract}

We have carried out a study of the chemical abundances of $^1$H,
$^4$He, $^{12}$C, $^{13}$C, $^{14}$N, $^{15}$N, $^{16}$O, $^{17}$O,
$^{20}$Ne and $^{22}$Ne in symbiotic stars (SSs) by means of a
population synthesis code. We find that the ratios of the number of
O-rich SSs to that of C-rich SSs in our simulations are between 3.4
and 24.1, depending on the third dredge-up efficiency $\lambda$ and
the terminal velocity of the stellar wind $v(\infty)$. The fraction
of SSs with $extrinsic$ C-rich cool giants in C-rich cool giants
ranges from 2.1\% to 22.7\%, depending on $\lambda$, the common
envelope algorithm and the mass-loss rate. Compared with the
observations, the distributions of the relative abundances of
$^{12}$C/$^{13}$C vs. [C/H] of the cool giants in SSs suggest that
the thermohaline mixing in low-mass stars may exist. The
distributions of the relative abundances of C/N vs. O/N, Ne/O vs.
N/O and He/H vs. N/O in the symbiotic nebulae indicate that it is
quite common that the nebular chemical abundances in SSs are
modified by the ejected materials from the hot components. Helium
overabundance in some symbiotic nebulae may be relevant to a helium
layer on the surfaces of white dwarf accretors.
\end{abstract}


\section{Introduction}
Symbiotic stars (SSs) are usually interacting binaries, composed of
a cool star, a hot component and a nebula. The cool component is a
red giant which is a first giant branch (FGB) or an asymptotic giant
branch (AGB) star. The hot component is a white dwarf (WD), a
subdwarf, an accreting low-mass main-sequence star, or a neutron
star \citep{kw84,m91,y95,it96}. In general, SSs are low-mass
binaries with an evolved giant transferring materials to a hot WD
companion which burns the accreted materials more or less steadily
\citep{m07}. According to the observed characteristics of their
specular spectra and photometries, SSs may stay in either the
quiescent phase or the outburst phase. During the quiescent phase,
SSs are undergoing a stable hydrogen burning on the surfaces of the
WD accretors. The outbursts may be due to either the thermonuclear
outbursts on the surfaces of accreting WDs which are called as
symbiotic novae \citep{ty76,pr80}, or accretion-disk instabilities
around a nearly steady burning WDs which are called as multiple
outbursts \citep{m05,m07}. Recent reviews of the properties of SSs
can be found in \citet{ms99} and \citet{m03,m07}.


The cool components usually have a high mass-loss rate in the SSs
with WD accretors. Their stellar winds have the chemical abundances
of the red giants. Parts of the stellar winds are ionized by the hot
components. The abundance determinations with nebular diagnostic
tools are possible \citep{s92}. SSs provide a good opportunity for
measuring the abundances of the red giants. The hot components may
support an additional high-velocity wind during the symbiotic nova
outbursts \citep{kw84}. High-velocity outflows have been observed in
very broad emission lines in essentially all the symbiotic novae
\citep{kt05}. \cite{lt94} analyzed classical nova abundances and
concluded that enhanced concentrations of heavy elements are
significant. Therefore, the abundances of the ejected materials from
the hot components in SSs may be different from those of the cool
giant stellar winds. However they are similar to those of the
classical novae. In general, the hot components have high luminosity
and effective temperature (They are similar to the central stars of
planetary nebulae. See Figure 5 in \citealt{m91}) so that they can
ionize the nebulae. In some eruptive SSs, the nebulae can also be
ionized by the region where the winds from the hot and cold
components collide \citep{w84}. It is difficult to determine the
origin of the nebulae in SSs. Based on emission line fluxes from
C$_{\rm III}$, C$_{\rm IV}$, N$_{\rm III}$, N$_{\rm IV}$ and O$_{\rm
III}$, \cite{n88} found that the CNO abundance ratios of the nebulae
in SSs are in the transition region from giants to supergiants and
concluded that the nebula is mainly due to mass lost by the red
giant. On the other hand, \cite{nv89} suggested that the flux ratios
of the emission line depend on the relative abundances of the two
winds in their study on Z And. \cite{vn92} found that CNO abundance
ratios of PU Vul are of the characteristic of novae.

In short, SSs offer an exciting laboratory for studying novae, the
red giants and the interaction of the two winds. The chemical
abundances of the ejected materials from the hot components, the
stellar wind of the cool components and the symbiotic nebulae are
the key factor to understand them. Up to now, a series of
observational data of the chemical abundances in SSs have been
published. \citet{n88}, \citet{dc92}, \citet{cd94} and \citet{lc05}
gave the chemical abundances of some symbiotic nebulae. \citet{s92}
and \citet{s06} showed the chemical abundances of the cool giants in
several SSs. However, there are few theoretical studies about it.
Recently, \citet{kp97} and \citet{jh98} carried out detailed
numerical simulations for the chemical abundances of the novae with
CO WD or ONe WD accretors. This makes it possible to simulate the
chemical abundances of the symbiotic novae. \cite{gj93},
\cite{WG98}, \cite{k02}, \cite{i04}, \cite{I04} and \cite{mg07}
developed thermally pulsing asymptotic giant branch (TP-AGB)
synthesis by which the chemical abundances of the cool components
can be computed. \citet{lyh06} (hereafter Paper I) constructed the
models for the SSs' populations. According to the above results or
models, it is possible to set up a preliminary model for theoretical
study on the chemical abundances of SSs.

In the present paper, the study of the chemical abundances of SSs
are carried out by means of a population synthesis code. In $\S$ 2
we present our assumptions and describe some details of the modeling
algorithm. In $\S$ 3 we discuss the main results and the effects of
different parameters. $\S$ 4 contains the main conclusions.

\section{Model}
For the binary evolution, we use a rapid binary star evolution (BSE)
code of \cite{h02}. Below we describe our algorithm from several
aspects.

\subsection{Symbiotic Stars}
Paper I carried out a detailed study of SSs. Comparing the observed
distributions with the those predicted of the orbital periods and
the hot component masses in SSs, it is worth to mention that there
is remarkable disagreement between the observations and predictions.

\cite{m07} showed the distributions of the measured orbital period
in about 70 SSs. The vast majority ($\sim 75\%$) of these systems
have orbital periods shorter than 1000 days while Paper I predicted
the distribution of longer orbital periods. The main reasons results
from two aspects:
\begin{enumerate}
\item It is hard to measure long orbital periods. All the SSs
with measured orbital periods in \cite{b00} and \cite{m03} are
S-type and about half of them are eclipsing binaries. Considering
the amplitude of radial velocity changes or eclipses, it is easier
to measure the short orbital periods. About 15 percent of 27 SSs in
\cite{m03} have longer orbital periods than 1000 days while it is
about 25 percent in \cite{m07}. There would be more SSs with long
orbital periods with further detailed observations. \item In Paper
I, SSs are detached binary systems and the process of mass transfer
results from the stellar wind of cool giants. To our knowledge, if
the mass ratio of the components ($q=M_{\rm donor}/M_{\rm
accretor}$) at onset of Roche lobe overflow (RLOF) is larger than a
certain critical value $q_{\rm c}$, the mass transfer is dynamically
unstable and results in the formation of a common envelop in short
time scale (about 100 years). The issue of the criterion for
dynamically unstable RLOF $q_{\rm c}$ is still open.
\citet{han01,han02} showed that $q_{\rm c}$ depends heavily on the
assumed mass-transfer efficiency. Paper I adopted $q_{\rm c}$ after
\cite{h02} in which the mass-transfer efficiency is 1. In Paper I,
$q_{\rm c}$ is usually smaller than 1.0. The observed $M_{\rm
giant}/M_{\rm WD}$s in SSs are between 2.0 and 4.0 and the predicted
those in Paper I are from 1.0 to 4.0, which means that there should
not be stable RLOF in SSs. However, \cite{m07} suggested that RLOF
can be quite common in symbiotic binaries with orbital periods
shorter than 1000 days. If this were true, the theoretical model of
mass transfer should be advanced. But this is out of the scope of
this paper.
\end{enumerate} The disagreement between the observed and the
predicted distributions of orbital periods in Paper I resulted from
observational biases and poor knowledge of the mass transfer
mechanism.

\cite{m07} showed that most of SSs have WD masses less than 0.6
$M_\odot$, while Paper I predicted the distribution of higher WD
masses. The disagreement is due to the following:
\begin{enumerate}
\item For mass estimates, an orbital inclination of $i=90^0$
or a limit to $i$ (see table 2 in \citet{m03}) is typically assumed,
hence the estimates are lower limits. \item Paper I used the
mass-loss law suggested by \cite{vw93} for AGB stars. However,
\cite{m07} suggested that both the symbiotic giants and Miras have
higher mass-loss rates than single giants or field Miras,
respectively. Therefore, Paper I may have overestimated the hot
component masses. \end{enumerate}

In short, due to poor knowledge of the mass loss from the giants and
the mass transfer mechanism in SSs, Paper I can only crudely predict
some observed characteristics. In this work, we accept still all the
criterions and the concepts on SSs in it. Following Paper I, we
assume that binary systems are considered as SSs if they satisfy the
following conditions:
\begin{itemize}
\item The systems are detached. \item The luminosity of the hot
component is higher than 10$L_\odot$ which is the `threshold'
luminosity for the hot component of SSs as inferred by \citet{m91}
and \citet{mk92}. This may be due to the thermonuclear burning
(including novae outbursts, stationary burning and post-eruption
burning). \item The hot component is a WD and the cool component is
a FGB or an AGB star.
\end{itemize} The liberation of gravitational energy by the accreted
matter may make the luminosity of the component larger than
10$L_\odot$. A detailed accretion model of SSs was discussed in
Paper I. Here, we do not model it.

All symbiotic phenomena in our work are produced by the hydrogen
burning on the WD surface. All SSs should stay in one of the
following three phases:
\begin{itemize}
\item the stable hydrogen burning phase;
\item the thermonuclear outburst phase (symbiotic nova);
\item the declining phase after a thermonuclear outburst.
\end{itemize} Paper I and the present paper do not
simulate the accretion disk in SSs. As mentioned in the
Introduction, the multiple outbursts due to the accretion-disk
instabilities usually occur around nearly steady burning WDs
\citep{m07}. In this work, they can be included in the quiescent SSs
which are undergoing the stable hydrogen burning.


SSs are complex binary systems. There are many uncertain physical
parameters which can affect the population of SSs. Paper I showed
that the numbers of SSs and the occurrences of symbiotic novae are
greatly affected by the algorithm of common envelope evolution, the
terminal velocity of stellar wind $v(\infty)$ and the critical
ignition mass $\Delta M_{\rm crit}^{\rm WD}$ which is necessary mass
accreted by WDs for a thermonuclear runaway. The structure factor of
the stellar wind velocity $\alpha_{\rm W}$ and an optically thick
wind give a small uncertainty. In this work, we use the models of
SSs in Paper I and discuss the effects of the algorithm of common
envelope evolution, $v(\infty)$ and $\Delta M_{\rm crit}^{\rm WD}$
on the chemical abundances of SSs. Other
subordinate parameters are the same as in the standard model of Paper I.\\

Common Envelope:\ For the common envelope evolution, it is generally
assumed that the orbital energy of the binary is used to expel the
envelope of the donor with an efficiency $\alpha_{\rm ce}$:
\begin{equation}
 E_{\rm bind}=\alpha_{\rm ce}\Delta E_{\rm orb},
 \label{eq:alpha}
\end{equation}
where $E_{\rm bind}$ is the total binding energy of the envelope and
$\Delta E_{\rm orb}$ is the orbital energy released in the
spiral-in. \citet{n20} suggested to describe the variation of the
separation of components in the common envelopes by an algorithm.
This algorithm is founded on the equation for the system orbital
angular momentum balance which implicitly assumes the conservation
of energy:
\begin{equation}
\frac{\Delta J}{J}=\gamma\frac{M_{\rm e}}{M+m}, \label{eq:gamma}
\end{equation}
where $J$ is the total angular momentum and $\Delta J$ is the change
of the total angular momentum during common envelope phase. In the
above formula, $M$ and $M_{\rm e}$ are respectively the mass and the
envelope mass of the donor, and $m$ is the companion mass. Following
\cite{nt05} and Paper I, we call the formalism of Eq.
(\ref{eq:alpha}) $\alpha$-algorithm and that of Eq. (\ref{eq:gamma})
$\gamma$-algorithm, which are respectively simulated in different
cases (see Table \ref{tab:case}). We take the `combined' parameter
$\alpha_{\rm ce}\lambda_{\rm ce}$ as 0.5 for $\alpha$-algorithm.
$\lambda_{\rm ce}$ is a structure parameter that depends on the
evolutionary stage of the donor.
For $\gamma$-algorithm, $\gamma=1.75$. \\

$v({\infty})$:\ It is difficult to determine the terminal velocity
of stellar wind $v({\infty})$. \cite{w03} fitted the relation
between the mass-loss rates and the terminal wind velocities derived
from their CO(2-1) observation by
\begin{equation}
\log_{10} (\dot{M}/M_\odot{\rm yr}^{-1})=-7.40+\frac{4}{3}\log_{10}
(v({\infty})/{\rm km \, s^{-1}}). \label{eq:winters}
\end{equation}
The mass-loss rate is given by the formulation of \cite{vw93} or
\cite{b95} and $v(\infty)$ can be obtained by Eq.
(\ref{eq:winters}). However, Eq. (\ref{eq:winters}) is valid for
$\dot{M}$ close to $ 10^{-6}M_\odot$ yr$^{-1}$. For a mass-loss rate
higher than $ 10^{-6}M_\odot$ yr$^{-1}$, Eq. (\ref{eq:winters})
gives too high $v(\infty)$. Based on the models of \citet{win00}, we
assume $v(\infty)=30 {\rm km\, s^{-1}}$ if $v(\infty)\geq30 {\rm
km\, s^{-1}}$. In the standard model of Paper I,
$v({\infty})=\frac{1}{2}v_{\rm esc}$ where $v_{\rm esc}$ is the
escape velocity. In the present paper, we also carry
out various $v(\infty)$s simulations.\\

$\Delta M_{\rm crit}^{\rm WD}$:\ The critical ignition mass of the
novae depends mainly on the mass of accreting WD, its temperature
and material accreting rate. \cite{y95} gave the `constant pressure'
expression for $\Delta M_{\rm crit}^{\rm WD}$ as
\begin{equation}
\frac{\Delta M_{{\rm crit}}^{{\rm WD}}}{M_\odot}=2\times
10^{-6}\left(\frac{M_{{\rm WD}}}{R^4_{{\rm WD}}}\right)^{-0.8},
\label{eq:yunm}
\end{equation}
where $M_{\rm WD}$ is the mass of WD accretor and $R_{{\rm WD}}$ is
the radius of zero-temperature degenerate objects \citep{n72},
\begin{equation}
R_{{\rm WD}}=0.0112R_\odot[(M_{{\rm WD}}/M_{{\rm
ch}})^{-2/3}-(M_{{\rm WD}}/M_{{\rm ch}})^{2/3}]^{1/2},
\end{equation}
where $M_{{\rm ch}}=1.433M_{\odot}$ and $R_{\odot}=7\times 10^{10}$
cm. \citet{n04} gave numerical fits to the critical ignition masses
for novae models calculated by \citet{pk95}. In most simulations, we
adopt Eq. (\ref{eq:yunm}) for $\Delta M_{\rm crit}^{\rm WD}$.
However, in order to investigate the influences of the $\Delta
M_{\rm crit}^{\rm WD}$ on our results, we also carry out a
simulation for Eq. (A1) of \citet{n04} in which $\Delta M_{\rm
crit}^{\rm WD}$ depends on the mass accretion rates and masses of WD
accretors.


\subsection{Chemical Abundances on the Surface of the Giant Stars}
\label{sec:abun} For a single star, three dredge-up processes and
hot bottom burning in a star with initial mass higher than
$4M_\odot$ may change the chemical abundances of the stellar
surface. We accept the prescriptions of \S 3.1.2 and 3.1.3 in
\cite{i04} for the first dredge-up during the first giant branch and
the second dredge-up during early asymptotic giant branch (E-AGB).
For the third dredge-up (TDU) and the hot bottom burning during the
TP-AGB phase, we use the TP-AGB synthesis in \cite{gj93},
\cite{k02}, \cite{i04}, \cite{I04} and \cite{mg07}. All details can
be found in Appendix A. In the present paper, we give the chemical
evolutions of $^1$H, $^4$He, $^{12}$C, $^{13}$C, $^{14}$N, $^{15}$N,
$^{16}$O, $^{17}$O, $^{20}$Ne and $^{22}$Ne on the stellar surface.

SSs in our work are binary systems. The binary mass transfer can
change the chemical abundances of the stellar surface. In binary
systems, there are two ways to transfer mass: (i)accretion from the
stellar wind material of a companion; (ii)Roche lobe overflow. BSE
model contains a standard Bondi-Hoyle type wind accretion
\citep{bh44} and a conservative mass transfer during the stable
Roche lobe overflow. After a star obtains $\Delta M$ from its
companion, the chemical abundances  of the stellar surface $X_2$ is
given by
\begin{equation}
X^{\rm new}_2=\frac{X^{\rm old}_2\times M^{\rm
env}_2+X_1\times\Delta M}{M^{\rm env}_2+\Delta M},
\end{equation}
where $M^{\rm env}$ is the envelope mass and $X_1$ is the chemical
abundances of its companion.
\subsection{Chemical Abundances of the Ejected Materials from the Hot WD Accretors}
The hot WD accretors eject materials from their surfaces.
\cite{kp97} and \cite{jh98} showed detailed study on their chemical
abundances during the thermonuclear outbursts. There is no direct
observational evidence that SSs in the stable hydrogen burning phase
and the declining phase lead to winds from the hot components.
Theoretically, the hot companions have high luminosity and
temperature during the above two phases. Some parts of the materials
on the surface of the accreting WDs may be blown off due to high
luminosity \citep{it96}. To our knowledge, no literature refers to
them in detail yet. However, their chemical abundances are possibly
between those of the ejected materials during the thermonuclear
outbursts and the stellar winds from the giant stars. In the present
paper, we only give the chemical abundances of the ejected materials
during the thermonuclear outbursts.

Based on the typical characteristics of the observed systems, nova
outbursts are usually divided into symbiotic novae and classical
novae. The classical novae can only last several days or weeks and
the variation of their visual magnitudes are between -5 and -10. The
symbiotic novae usually last several decades and the variation of
their visual magnitudes are between -3 and -8. Thermonuclear
runaways appear to be the most promising mechanism for classical
novae \citep{k86}. \cite{ty76} suggested that the above mechanism is
applicable to symbiotic novae. The observational differences between
the symbiotic novae and the classical novae may stem from the
properties of the binary systems \citep{it96}. The physical nature
of the classical novae and the symbiotic novae is identical.
Therefore, we assume that the thermonuclear outbursts occurring in
our binary systems selected for SSs are the symbiotic novae.

Numerical studies of the nova outbursts have been carried out by
\citet{spk93,kp94,pk95,y05}. \cite{kp97} published detailed
multicycle calculations of the nova outbursts for CO WDs with mass
ranging from 0.65 to 1.4 $M_\odot$. In their simulations, the
element abundances of nova ejecta are affected by the four basic and
independent parameters: C/O ratio in the accreting WD, its core
temperature $T_{\rm WD}$, the mass accreting rate $\dot {M}_{\rm
WD}$ and the mass $M_{\rm WD}$. In order to test the effect of the
WD composition on the abundances of nova ejecta, \cite{kp97}
calculated the models with the accreting WDs composed of pure-C,
pure-O and C/O=1, respectively. They found that the WD composition
is not reflected in the abundances of ejecta. In an extensive study
of close binary evolution, \cite{it85} showed that the AGB phase may
be suppressed in a close binary and a WD formed in such a system
should have a ratio very close to unity. In our work, we assume that
the CO WD composition is C/O=1. A WD temperature of $10^7$ K
corresponds to an age of $10^9$ yr \citep{it84}. In fact, for most
of SSs, the time interval between the formation of the WD and the
beginning of the symbiotic stage may be longer than it, up to
$10^{10}$ yr \citep{y95}. Neglecting the effect of the nova
outbursts on the temperature of WD accretor $T_{\rm WD}$, we assume
that it is $10^7$ K. We select 10 models in \cite{kp97} in which
$T_{\rm WD}=10^7$ K and C/O ratio of the accreting WD is 1. Their
element abundances are determined by the mass accreting rate $\dot
{M}_{\rm WD}$ and $M_{\rm WD}$. By a bilinear interpolation
\citep{p92} of 10 models in \cite{kp97}, the abundances of $^1$H,
$^4$He, $^{12}$C, $^{13}$C, $^{14}$N, $^{15}$N, $^{16}$O and
$^{17}$O in the nova ejecta are calculated. If $\dot {M}_{\rm WD}$
or $M_{\rm WD}$ in SSs are not in the range of the bilinear
interpolation, they are taken as the most vicinal those in the 10
models.

\cite{kp97} did not give the abundances of $^{20}$Ne and $^{22}$Ne.
\cite{jh98} gave the nucleosynthesis in the nova outbursts with CO
and ONe WDs. In their calculations, the nova nucleosynthesis are
affected by $M_{\rm WD}$, $\dot {M}_{\rm WD}$, the initial
luminosity (or $T_{\rm WD}$) and the degree of mixing between core
and envelope. To test the effect of the WD mass, they carried out a
number of simulations involving both CO WD ($M_{\rm WD}$=0.8, 1.0
and 1.15 $M_\odot$) and ONe ones ($M_{\rm WD}$=1.0, 1.15, 1.25 and
1.35$M_\odot$). $\dot {M}_{\rm WD}$ is $2\times10^{-10} M_\odot$
yr$^{-1}$ and their initial luminosity is $10^{-2} L_\odot$.  The
degree of mixing between the core and the envelope is a very
uncertain parameter. \cite{jh98} modeled three different mixing
levels: 25\%, 50\% and 75\%. Their results showed that higher mixing
degree favors the synthesis of higher metal nuclei in ONe WDs.
Following \cite{s98}, we adopt a 50\% degree of the mixing. For the
simulations of CO WDs in \cite{jh98}, we select the three nova
models (The degree of the mixing is 50\%) to calculate the
abundances of $^{20}$Ne and $^{22}$Ne by the fitting formulae:
\begin{equation}
\begin{array}{ll}
\log X({^{20}\rm Ne})=&-3.206+0.14476M_{\rm WD},\\
\log X({^{22}\rm Ne})=&-2.57118+0.60784M_{\rm WD}-0.33769M_{\rm
WD}^2,\\
\end{array}
\label{eq:kp97}
\end{equation}
where $M_{\rm WD}$ is in solar unit and the formulae agree with the
numerical results to within a factor of 1.1. In the above fits, the
abundances of $^{20}$Ne and $^{22}$Ne depend weakly on $M_{\rm WD}$.
The range of CO WD mass $M_{\rm WD}$ is from $\sim$ 0.5 to 1.4
$M_\odot$. The abundances of $^{20}$Ne calculated by Eq.
(\ref{eq:kp97}) are between $\sim$ $10^{-3.13}$ and $10^{-3.00}$,
and the abundances of $^{22}$Ne are between $\sim$ $10^{-3.35}$ and
$10^{-3.38}$. Therefore, we use Eq. (\ref{eq:kp97}) to calculate
$X({^{20}\rm Ne})$ and $X({^{22}\rm Ne})$ of all symbiotic novae
with the CO WDs. For the nova of ONe WD, we fit data of Table 3 in
\cite{jh98} for a 50\% degree of mixing by:
\begin{equation}
\begin{array}{ll}
\log X(^1{\rm H})=&-1.486+1.982M_{\rm WD}-0.992M_{\rm WD}^2,\\
\log X(^4{\rm He})=&-0.839-0.103M_{\rm WD}+0.197M_{\rm WD}^2,\\
\log X(^{12}{\rm C})=&-10.664+14.798M_{\rm WD}-6.025M_{\rm WD}^2,\\
\log X(^{13}{\rm C})=&-14.691+22.517M_{\rm WD}-9.607M_{\rm WD}^2,\\
\log X(^{14}{\rm N})=&4.504-11.174M_{\rm WD}+5.079M_{\rm WD}^2,\\
\log X(^{15}{\rm N})=&-2.196-2.313M_{\rm WD}+2.404M_{\rm WD}^2,\\
\log X(^{16}{\rm O})=&-9.835+17.609M_{\rm WD}-8.550M_{\rm WD}^2,\\
\log X(^{17}{\rm O})=&-14.492+22.095M_{\rm WD}-9.366M_{\rm WD}^2,\\
\log X(^{20}{\rm Ne})=&-0.536-0.192M_{\rm WD},\\
\log X(^{22}{\rm Ne})=&-21.20356+34.46454M_{\rm WD}-15.9763M_{\rm
WD}^2,\\
\end{array}
\label{eq:jh98}
\end{equation}
which agree with the numerical results to within a factor of 1.3.

We neglect other elements because their abundances are much smaller
than the above elements or their isotopes. All abundances in the
nova ejecta are renormalized so that their sum is 1.0.

\subsection{Symbiotic Nebulae}
\label{sec:syne}

In SSs, there may be two winds. One comes from the cool giant and
its chemical abundances are similar to those of the red giant
envelope and its velocity is between $\sim$ 5 and 30 km s$^{-1}$.
The other may come from the hot components. The components during
the symbiotic outbursts have the winds with a high velocity ($\sim$
1000 km s$^{-1}$) although there is not a detailed description on
the wind from the hot components during the quiescent phase and the
declining phase \citep{k86}. Wind collision is inevitable in the
symbiotic outbursts. Recently, \cite{kt05,kt07} carried out a
detailed study for the colliding winds in SSs. According to their
simulations, the structure of the colliding winds (including
temperature and density) is very complicated. Symbiotic nebulae
should have a similar structure with the colliding winds at least
during the symbiotic outbursts.

In order to avoid the difficulty of modeling the real symbiotic
nebulae and study the nebular chemical abundances by population
synthesis method, we assume roughly that the compositions of the
symbiotic nebulae undergo two independent phases:
\begin{itemize}
\item The symbiotic nebula is mainly composed of the ejected materials
from the hot WD when the thermonuclear runaway occurs and this phase
last for $t_{\rm on}$. The lasting time scale $t_{\rm on}$ on which
the hot components are in a `plateau' phase with high luminosity is
given by Eq. (30) of Paper I. At this phase, the symbiotic nebula
embodies the chemical characteristics of a nova. \item After $t_{\rm
on}$, the symbiotic nebula is mainly composed of the stellar wind
materials from the cool giants until the next symbiotic nova occurs.
During this phase, the symbiotic nebula embodies the chemical
characteristics of the cool giants in our models. For stable
hydrogen burning SSs, the symbiotic nebula is always composed of the
stellar wind materials from the cool giants.
\end{itemize}

However, some symbiotic novae have helium WD accretors. To our
knowledge, no literature refers to their chemical abundance.
According to Paper I, the occurrence rate of the symbiotic novae
with helium WDs is at most about $\frac{1}{75}$ of that with CO and
ONe WDs. We do not consider SSs with helim WD accretors in this
paper.

\subsection{Basic Parameters of the Monte Carlo Simulation }
We carry out binary population synthesis via Monte Carlo simulation
technique in order to obtain the properties of SSs' population. For
the population synthesis of binary stars, the main input model
parameters are: (i) the initial mass function (IMF) of the
primaries; (ii) the mass-ratio distribution of the binaries; (iii)
 the distribution of orbital separations; (iv) the eccentricity
distribution; (v) the metallicity $Z$ of the binary systems.

We use a simple approximation to the IMF of \citet{ms79}. The
primary mass is generated using  the formula suggested by
\citet{e89}
\begin{equation}
M_1=\frac{0.19X}{(1-X)^{0.75}+0.032(1-X)^{0.25}},
\end{equation}
where $X$ is a random variable uniformly distributed in the range
[0,1],  and $M_1$ is the primary mass from $0.8M_\odot$ to
$8M_\odot$.

For the mass-ratio distribution of binary systems, we consider only
a constant distribution \citep{m92,gm94},
\begin{equation}
n(q)=1,~~    0< q \leq 1,
\end{equation}
where $q=M_2/M_1$.

The distribution of separations is given by
\begin{equation}
\log a =5X+1,
\end{equation}
where $X$ is a random variable uniformly distributed in the range
[0,1] and $a$ is in $R_\odot$.

In our work, the metallicity $Z$=0.02 is adopted. We assume that all
binaries have initially circular orbits, and we follow the evolution
of both components by BSE code, including the effect of tides on
binary evolution \citep{h02}. We take $2\times10^5$ initial binary
systems for each simulation. Since we present, for every simulation,
the results of one run of the code, the numbers given are subject to
Poisson noise. For simulations with $2\times10^5$ binaries, the
relative errors of the numbers of the symbiotic systems in different
simulations are lower than 5\%. Thus, $2\times10^5$ initial binaries
appear to be an acceptable sample for our study.

We assume that one binary with $M_1\geq 0.8 M_\odot $ is formed
annually in the Galaxy to calculate the birthrate of SSs
\citep{y94,h95a,h95b}.

\begin{table*}
 \begin{minipage}{170mm}
  \caption{Parameters of the models of the population of SSs. The first column
           gives the model number. Columns 2, 3, 4 and 5 show the mass-loss rate $\dot{M}$,
           TDU efficiency $\lambda$, the minimum core mass for TDU
           and the inter-shell abundance, respectively.
           The detailed descriptions on the above parameters are given in Appendix A.
           The 6th 7th and 8th columns show the algorithm of the common envelope,
           the terminal velocity and the critical ignition mass, respectively.
           N04 in column 8 means \citet{n04}.
           }
  \tabcolsep0.8mm
  \begin{tabular}{llllllll}
  \hline
Cases & $\dot{M}$&$\lambda$& $M_{\rm c}^{\rm min}$&Inter-Shell Abundances&Common Envelope&$v(\infty)$&$\Delta M_{\rm crit}^{\rm WD}$ \\
case 1& Eq.(\ref{eq:vwml})&Eq.(\ref{eq:lamb})&Eq.(\ref{eq:mcmin})      & \cite{mg07}&$\alpha_{\rm ce}\lambda_{\rm ce}=0.5$&Eq.(\ref{eq:winters})   &Eq.(\ref{eq:yunm})\\
case 2& Eq.(\ref{eq:bml}) &Eq.(\ref{eq:lamb})&Eq.(\ref{eq:mcmin})      & \cite{mg07}&$\alpha_{\rm ce}\lambda_{\rm ce}=0.5$&Eq.(\ref{eq:winters})   &Eq.(\ref{eq:yunm})\\
case 3& Eq.(\ref{eq:vwml})&0.5               &Eq.(\ref{eq:mcmin})      & \cite{mg07}&$\alpha_{\rm ce}\lambda_{\rm ce}=0.5$&Eq.(\ref{eq:winters})   &Eq.(\ref{eq:yunm})\\
case 4& Eq.(\ref{eq:vwml})&0.75              &Eq.(\ref{eq:mcmin})      & \cite{mg07}&$\alpha_{\rm ce}\lambda_{\rm ce}=0.5$&Eq.(\ref{eq:winters})   &Eq.(\ref{eq:yunm})\\
case 5& Eq.(\ref{eq:vwml})&Eq.(\ref{eq:lamb})&0.58$M_\odot$            & \cite{mg07}&$\alpha_{\rm ce}\lambda_{\rm ce}=0.5$&Eq.(\ref{eq:winters})   &Eq.(\ref{eq:yunm})\\
case 6& Eq.(\ref{eq:vwml})&Eq.(\ref{eq:lamb})&Eq.(\ref{eq:mcmin})      & \cite{i04} &$\alpha_{\rm ce}\lambda_{\rm ce}=0.5$&Eq.(\ref{eq:winters})   &Eq.(\ref{eq:yunm})\\
case 7& Eq.(\ref{eq:vwml})&Eq.(\ref{eq:lamb})&Eq.(\ref{eq:mcmin})      & \cite{mg07}&$\gamma=1.75$                        &Eq.(\ref{eq:winters})   &Eq.(\ref{eq:yunm})\\
case 8& Eq.(\ref{eq:vwml})&Eq.(\ref{eq:lamb})&Eq.(\ref{eq:mcmin})      & \cite{mg07}&$\alpha_{\rm ce}\lambda_{\rm ce}=0.5$&$\frac{1}{2}v_{\rm esc}$&Eq.(\ref{eq:yunm})\\
case 9& Eq.(\ref{eq:vwml})&Eq.(\ref{eq:lamb})&Eq.(\ref{eq:mcmin})      & \cite{mg07}&$\alpha_{\rm ce}\lambda_{\rm ce}=0.5$&Eq.(\ref{eq:winters})   &Eq.(A1) of N04\\
\hline
 \label{tab:case}
\end{tabular}
\end{minipage}
\end{table*}
\section{Results}
We construct a set of models in which we vary different input
parameters relevant to the chemical abundances of SSs. Table
\ref{tab:case} lists all cases considered in this work. Many
observational evidences showed that the terminal velocity of stellar
wind $v(\infty)$ increases when a star ascends along the AGB
\citep{o02,w03,b05}. In this work, we take $v(\infty)$ calculated by
Eq. (\ref{eq:winters}) as the standard terminal velocity of stellar
wind. Case 1 is the standard model in this work. The results of SSs'
population are shown in Table \ref{tab:result}.
\begin{table*}
 \begin{minipage}{170mm}
  \caption{Different models of the SSs' population. The first column
           gives the model number according to Table \ref{tab:case}.
           Column 2 shows the birthrate of SSs in
           the Galaxy. From columns 3 to 5, the number of O-rich SSs,
           C-rich SSs and all SSs are given,
           respectively. In the 3rd and 4th columns, the numbers in
           parentheses mean the numbers of O-rich and C-rich SSs in
           which we neglect the effects of the mass transfer in
           binary systems on the chemical abundances on their
           surfaces, respectively.
           Columns 6 and 7 show the ratios of the number of SSs in the cooling
           phase and the stable hydrogen burning phase to their total number, respectively.
           Columns 8, 9 and 10 give the
           occurrence rates of SyNe (symbiotic novae) with the accreting CO WDs, ONe WDs and
           total rates. The number of SyNe with accreting CO
           and ONe WDs are shown in columns 11 and 12, respectively.
           }
 \tabcolsep0.1mm
  \begin{tabular}{ccccccccccccc}
  \hline
\multicolumn{1}{c}{Cases}&\multicolumn{1}{c}{Birthrate}&\multicolumn{3}{c}{Number
of  SSs}&\multicolumn{1}{c}{$N_{\rm
tcool}$}&\multicolumn{1}{c}{$N_{\rm
stable}$}&\multicolumn{3}{c}{Occurrence rate of
SyNe (yr$^{-1})$}&\multicolumn{2}{c}{Number of SyNe}\\
  & of SSs (yr$^{-1}$)&O-rich&C-rich&Total&$\overline{N_{\rm total}}$&$\overline{N_{\rm total}}$&CO & ONe &Total&CO&ONe\\
1&2&3&4&5&6&7&8&9&10&11&12\\
case 1& 0.119&4800 (4860)&870 (810)&5670& 0.57&0.25&3.8&0.2&4.0&1000&4\\
case 2& 0.118&5850 (6070)&1240(1020)&7100& 0.42&0.45&3.4&0.2&3.6&930 &3\\
case 3& 0.119&5340 (5390)&220 (170)&5560& 0.59&0.24&3.7&0.2&3.9&940 &3\\
case 4& 0.119&4930 (4970)&840 (800)&5770& 0.58&0.24&3.9&0.2&4.1&1000&4\\
case 5& 0.111&4330 (4410)&1220(1140)&5550& 0.56&0.26&3.8&0.2&4.0&1000&4\\
case 6& 0.119&4740 (4830&930  (840)&5660& 0.57&0.25&3.8&0.2&4.0&1000&4\\
case 7& 0.160&7900 (7930)&1440(1410)&9340& 0.49&0.33&5.3&0.4&5.7&1700&12\\
case 8& 0.072&2370 (2450)&700 (620)&3070& 0.28&0.60&1.1&0.1&1.2&370 &4\\
case 9& 0.115&7830 (7900)&1790(1720)&9620& 0.75&0.15&9.8&0.4&10.2&1000&4\\
\hline
 \label{tab:result}
\end{tabular}
\end{minipage}
\end{table*}

\begin{figure}
\includegraphics[totalheight=3.in,width=2.8in,angle=-90]{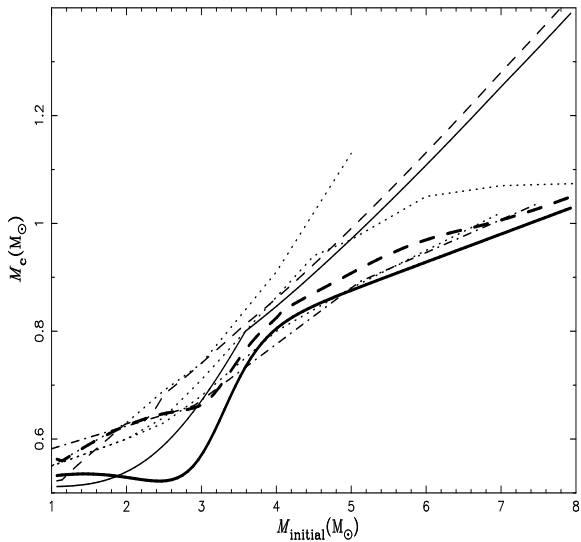}
\caption{Initial-final mass relations.  Thin and thick solid lines
        are the core masses at the first thermal pulse in \citet{h00} and
        in our work (See Eq. (\ref{eq:mc1tp})), respectively. Thin and thick dashed
        lines are the final masses in \citet{h00} and case 1 in this work, respectively.
        The dotted lines (from top to bottom) represent the relations from \citet{g00},
        \citet{h95} and \citet{w00}, respectively. The dot-dashed line represent the relation from \citet{h94}.}
\label{fig:mc}
\end{figure}

\subsection{Galactic Birthrate and the Number of SSs}
First, we discuss the gross properties of the modeled population of
SSs as those in Paper I. As Table \ref{tab:result} shows, the
Galactic birthrate of SSs may range from 0.072 (case 8) to 0.160
(case 7) yr$^{-1}$ and their number is from about 3070 (case 8) to
9620 (case 9). The occurrence rate of the symbiotic novae is between
1.2 (case 8) and 10.2 (case 9) yr$^{-1}$ and the number of the
symbiotic novae is from about 370 (case 8) to 1700 (case 7). The
contribution of the symbiotic novae with ONe WD accretors to the
occurrence rate is negligible in all cases, which is greatly
different from that in Paper I. The main reason is the different
stellar evolutions adopted during the TP-AGB phases.

All assumptions in cases 1 and 8 are respectively the same as those
in case 7 and the standard model of Paper I except the stellar
evolutions during the TP-AGB phase. The greatest difference of the
TP-AGB phase between this work and Paper I is the initial-final mass
relations. It is well known that the relation is still very
uncertain. Fig. \ref{fig:mc} shows the initial-final mass relations
in different literatures. Using the mass-loss rate of \cite{vw93}
and assuming the classical core mass luminosity relation, \cite{w00}
gave an initial-final mass relation which is plotted in Fig.
\ref{fig:mc}. The relation for the more massive stars in \cite{g00}
is steep because TDU and hot bottom burning were not included. By
applying new observational data (including NGC2168, NGC2516, NGC2287
and NGC3532) and improved theory (stellar evolution calculations
including new opacities and main sequence overshoot; the extensive
mass grid of WD cooling sequences including models with thick and
thin He- and H-layer), \citet{h95} gave an initial-final mass
relation which has a less slope than that in \cite{g00} for the more
massive stars. Based on new stellar evolution calculations
(including TDU and hot bottom burning) and new observational data,
\cite{w00} showed an initial-final mass relation which has lower
final mass than that of \citet{h95} in the more massive stars. Using
a criterion for envelope ejection in AGB or FGB stars, \cite{h94}
obtained an initial-final mass relation which is very similar to
that in \cite{w00}. Paper I accepted the description of \citet{h00}
on TP-AGB evolution. The thin solid line and the dashed line in Fig.
\ref{fig:mc} represent the core mass at the first thermal pulse
$M_{\rm c, 1tp}$ and the initial-final mass relations in
\citet{h00}, respectively. In this paper, we obtain the final mass
by the synthesis TP-AGB evolutions (Detailed descriptions can be
seen in Appendix A ). The $M_{\rm c, 1tp}$ (See Eq. (8) in
\citealt{k02}) and the final mass are shown by the thick solid and
dashed lines in Fig. \ref{fig:mc}, respectively.  The relation of
\citet{h00} is steeper than ours for $M_{\rm initial}> 4M_\odot$
(especially $M_{\rm initial}> 6M_\odot$) and gives a higher WD's
mass than ours. In Paper I, we assumed that the ONe WD originates
from a star with initial mass being between 6.1 $M_\odot$ and about
8 $ M_\odot$\citep{p98}. In this work, we still keep the above
assumption. The masses of ONe WDs in this work are much lower than
those in Paper I. A small mass of the accreting WD is unfavorable
for producing the symbiotic phenomenon. Compared with case 7 and the
standard model in Paper I, the birthrate, the number of SSs and the
symbiotic novae decrease in cases 1 and 8. Especially, the
occurrence rate of the symbiotic novae with accreting ONe WDs in
case 1 is $\sim$ 3\% of that in case 7 of Paper I, and in case 8 it
is $\sim$ 5\% of that in the standard model of Paper I. Therefore,
the initial-final mass relation has great effects on the SSs'
population and especially on the symbiotic novae.

Using the observational data in the last 25 years, \cite{it96}
estimated that the occurrence rate of symbiotic novae is $\sim$ 1
yr$^{-1}$ within a factor of 3. As Table. \ref{tab:result} shows,
the occurrence rate in case 1 is $\sim$ 4 and close to the
occurrence rate estimated by \cite{it96}. Under the same conditions,
case 7 in Paper I gave a higher occurrence rate of $\sim$ 11.9
yr$^{-1}$. Obviously, an steep initial-final mass relation like that
in \cite{h00} overestimates the occurrence rate of symbiotic novae.
This indicates that the initial-final mass relations from
\cite{h94}, \cite{w00} and the present paper are more reliable than
steep ones. However, due to the complicity of SSs, the above point
of view needs supports from further theoretical study and more
observational evidence.

\subsection{The Effects of Input Parameters}
\label{sec:Parameters} In this section, we discuss the effects of
the input physical parameters on the SSs' population and the ratio
of the number of C-rich SSs in which the cool companions are C-rich
giants to that of O-rich SSs in which the cool companions are O-rich
giants. In Paper I, we analyzed the effects of the physical
parameters ( the common envelope algorithm, the terminal velocity of
the stellar wind $v(\infty)$ and the critical ignition mass of the
novae $\Delta M_{\rm crit}^{\rm WD}$ ) on the SSs' population. They
have a great effect on the SSs' population in our simulations, which
is similar to those in Paper I. Table \ref{tab:result} shows, the
{\it common envelope algorithm} and $\Delta M_{\rm crit}^{\rm WD}$
have weak effects on the ratio of the number of O-rich to that of
C-rich SSs while the $v(\infty)$ in case 8 takes an uncertainty
within  a factor of about 1.6. With stars ascending along the AGB,
the $v(\infty)$ in case 8 decreases while more C-rich giants are
formed. A small $v(\infty)$ is favor for the formation of SSs.
Therefore, the ratio of O-rich to C-rich SSs in case 8 decreases.
The effects of other physical parameters
are following. \\
{\it Mass-loss rate}:
Compared with \cite{vw93}(see Eq. (\ref{eq:vwml})), the mass loss
rate of \cite{b95}(see Eq. (\ref{eq:bml})) greatly depends on
stellar luminosity. When the cool giants ascend along the AGB, the
luminosities gradually increase. Therefore, case 2 gives a higher
mass-loss rate in the later phase during the AGB, which can increase
the SSs' number of the stable hydrogen burning while decrease the
SSs' number of the thermonuclear runaways. In general, C-rich giants
are formed at a later phase of the AGB by the third dredge-up. The
mass-loss rate in \cite{b95}(See Eq. (\ref{eq:bml})) enhances the
number of SSs and the number ratio of C-rich to O-rich SSs but
decreases the occurrence rate of the symbiotic novae and their
number. However, the effect is
weak.\\
{\it TDU efficiency $\lambda$ }:\ Comparing cases 1, 3 and 4, we
find that $\lambda$ has a great effect on the ratio of O-rich to
C-rich SSs' population. The effect of $\lambda$ on stellar evolution
has mainly two aspects, {\it i.e.} the core mass evolution and the
changes of the chemical abundances in stellar envelope. The average
dredge-up efficiency in case 1 is higher than that in case 3
($\lambda=0.5$) and approximated to that in case 4 ($\lambda=0.75$).
In cases 1, 3 and 4, the SSs' populations are almost the same.
However, the number ratio of O-rich to C-rich SSs in case 3 is 4
times more than the ratio in cases 1 or 4.
\\
{\it The minimum core mass for TDU $M_{\rm c}^{\rm min}$}: \ For the
same initial masses, the $M_{\rm c}^{\rm min}$ of \cite{k02}(See Eq.
(\ref{eq:mcmin})) is higher than 0.58$M_\odot$. In case 1, a star
with an initial mass higher than $\sim$ 2.0$M_\odot$ can undergo the
third dredge-up, but only 1.5$M_\odot$ in case 5. Therefore, C-rich
SSs are more easily produced in case 5. Its effect is within a
factor about 1.5.
\\
{\it The inter-shell abundances}: \ As cases 1 and 6 in Table
\ref{tab:result} show, the two different inter-shell abundances have
no effect on the SSs' populations. In \cite{i04}, $^{16}$O abundance
of the inter-shell region is lower than that in \cite{mg07}. The
number of C-rich SSs in case 6 increases. However, the effect of the
inter-shell abundances on ratio of O-rich to C-rich SSs is very
weak.

In short, the mass-loss rate and the inter-shell abundances have
very weak effect on the SSs' population and the ratio of O-rich to
C-rich SSs. The TDU efficiency $\lambda$ has a very weak effect on
the SSs' population, while the number ratio of O-rich to C-rich SSs
have a great dependence on $\lambda$.


\subsection{C-rich SSs}

\cite{b00} presented a catalogue of SSs which includes 188 SSs as
well as 28 objects suspected of being SSs. According to the spectral
types of the cool components listed by the catalogue, one can find
out 5 C-rich SSs in 176 Galactic SSs and the number ratio of O-rich
to C-rich SSs is about 35. In our simulations, from cases 1 to 9,
the number ratios of O-rich to C-rich SSs are about 5.5, 4.9, 24.1,
5.8, 3.6, 5.2, 5.6, 3.4 and 4.3, respectively. Our results are lower
than the above observational ratio. In all cases, the ratio of case
3 with a low $\lambda$=0.5 is the closest to the ratio observed in
\cite{b00}. However, according to the typical TP-AGB synthesis
\citep{gj93,k02,i04,mg07}, $\lambda$=0.5 may be too low for the
solar-like stars. In the solar neighborhood, the number ratio of
M-type giants to C-type giants is $\sim$ 5 in \cite{g02}, which is
close to our results except for case 3. We consider that the small
fraction of the observed C-rich giants in SSs may result from two
respects: (i)The C-rich giants may have high terminal velocity of
the stellar wind $v(\infty)$. When we use the fit of $v(\infty)$ in
\cite{w03}, we should note that the sample of \cite{w03} has 65
sources but only two of them are C-rich giants. Eq.
(\ref{eq:winters}) may be unsuitable for the C-rich giants.
Comparing cases 1 and 8, one can find that $v(\infty)$ has a great
effect on the number ratio of O-rich to C-rich SSs. We may have
underestimated $v(\infty)$ of C-rich giants. (ii)The cool giants in
SSs have high mass-loss rates. In cases 1 and 2, we use the
mass-loss rates of \cite{vw93} and \cite{b95}, respectively. The
average mass-loss rate of the former is higher than that of the
later during TP-AGB when the TDU occurs. The higher the mass-loss
rate is, the more quickly the envelope mass decreases. According to
Eq. (\ref{eq:tip}) in the present paper, the interpulse period
$\tau_{\rm ip}$ increases with the envelope mass decreasing. A long
$\tau_{\rm ip}$ reduces the TDU progressive number and the TDU
efficiency $\lambda$. The large mass-loss rate is unfavorable for
the formation of the carbon stars, which can be seen by comparing
the number ratio of O-rich to C-rich SSs in case 1 with that in case
2. \cite{m07} suggested that both the symbiotic giants and Miras
have higher mass-loss rates than single giants or field Miras,
respectively. Therefore, the predicted number ratio of O-rich to
C-rich SSs lower than the observational value may result from our
underestimating the $v(\infty)$ of C-rich giants and the mass-loss
rates of the cool giants in SSs.

Based on the formation channels, C-rich stars are classed into two
types. The $intrinsic$ C-rich stars mean that the carbon observed in
the atmosphere of the TP-AGB stars results from the third drudge-up.
The $extrinsic$ C-rich stars are the giants polluted by carbon-rich
matter from the former TP-AGB companion. In this work, we consider
the effects of the mass transfer in binary systems on the chemical
abundances on their stellar surfaces (See \S \ref{sec:abun}). In
order to check the above effects, we also carry out the simulations
in which the mass transfer between the two components can not vary
the abundances on their surfaces. The results are shown in the
parentheses of columns 3 and 4 of Table. \ref{tab:result}. The
ratios of SSs with the $extrinsic$ C-rich giants to the total C-rich
SSs from case 1 to case 9 are 6.8\%, 17.7\%, 22.7\%, 4.7\%, 6.6\%,
9.7\%, 2.08\%, 11.4\% and 3.9\%, respectively. The ratio is
sensitive to the efficiency of TDU $\lambda$, the common envelope
algorithm and the mass-loss rate. According to the discussions in \S
\ref{sec:Parameters}, we know that a high mass-loss rate occurs in
case 2 when carbon abundance is increased by the TDU. The
$extrinsic$ C-rich giants in case 2 are formed more easy than those
in case 1. A low $\lambda$ in case 3 can prevent the formation of
C-rich stars including $intrinsic$ and $extrinsic$ C-rich stars. The
high ratio in case 3 results from a small number of C-rich stars.
Having undergone the $\alpha$-algorithm common envelope evolution,
the orbital period of binary system becomes to be several percent of
that before common envelope evolution. With $\gamma$-algorithm in
case 7, post-common-envelope systems are wider than with the
$\alpha$-algorithm in case 1 and this facilitates a symbiotic
phenomenon, allowing more stars to evolve further along the FGB or
AGB before the second RLOF (Paper I). In general,
post-common-envelope systems hardly form the $extrinsic$ C-rich
stars but can form $intrinsic$ C-rich stars. Therefore, the ratio of
SSs with the $extrinsic$ C-rich giants to the total C-rich SSs in
case 7 is very low. In most cases, the fraction of $extrinsic$
C-rich SSs in total C-rich SSs is lower than 10\%. This indicates
that the pollution of the carbon rich materials from the former
TP-AGB companions on the cool components is weak so that most of
C-rich giants in SSs are $intrinsic$ C-rich stars. \cite{m07} showed
that all C-rich giants observed in SSs are $intrinsic$ C-rich stars,
which is in agreement with our results.

\subsection{$^{12}$C/$^{13}$C vs. [C/H] of the Cool Giants in SSs}
\begin{figure*}
\begin{tabular}{c@{\hspace{3pc}}c}
\includegraphics[totalheight=3.2in,width=4in,angle=-90]{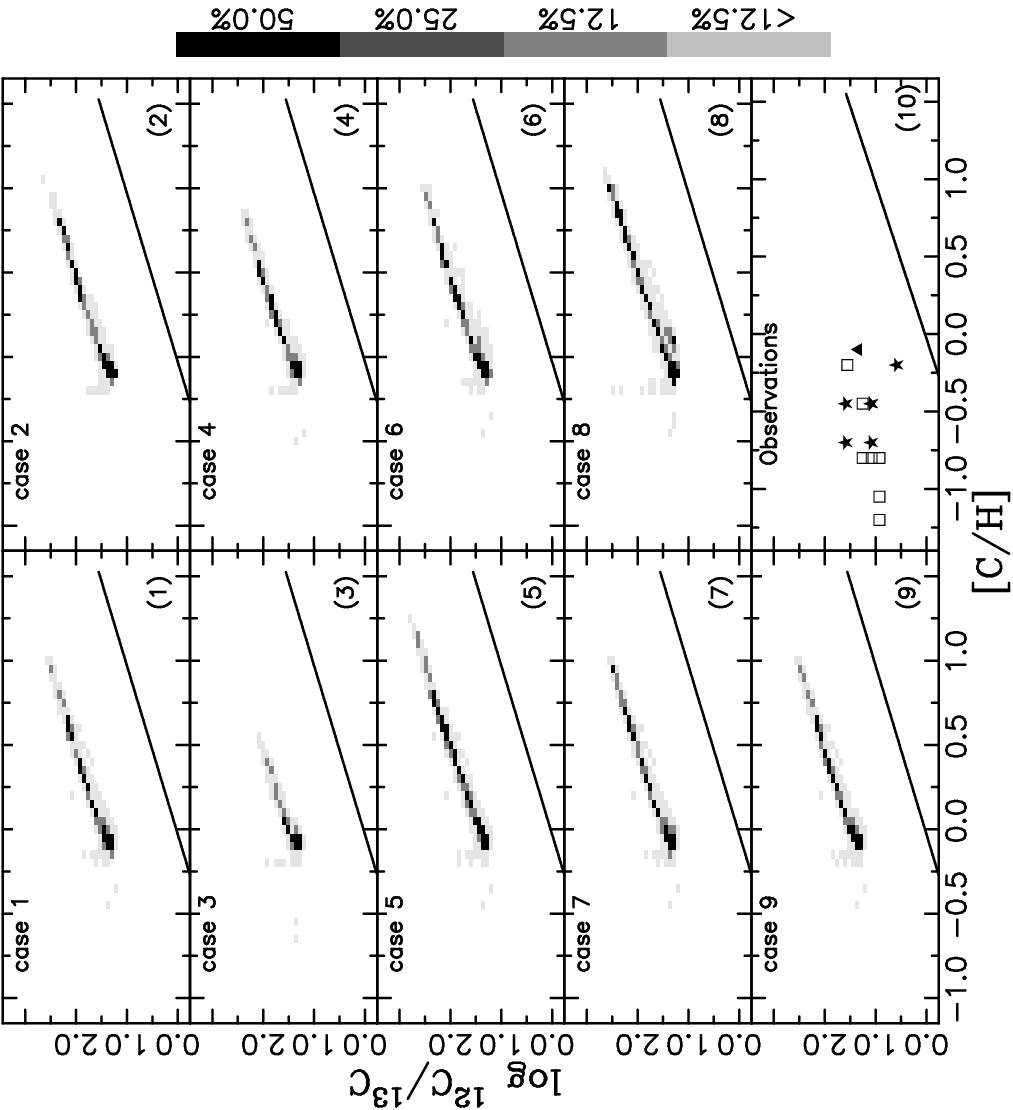}&
\includegraphics[totalheight=3.2in,width=4in,angle=-90]{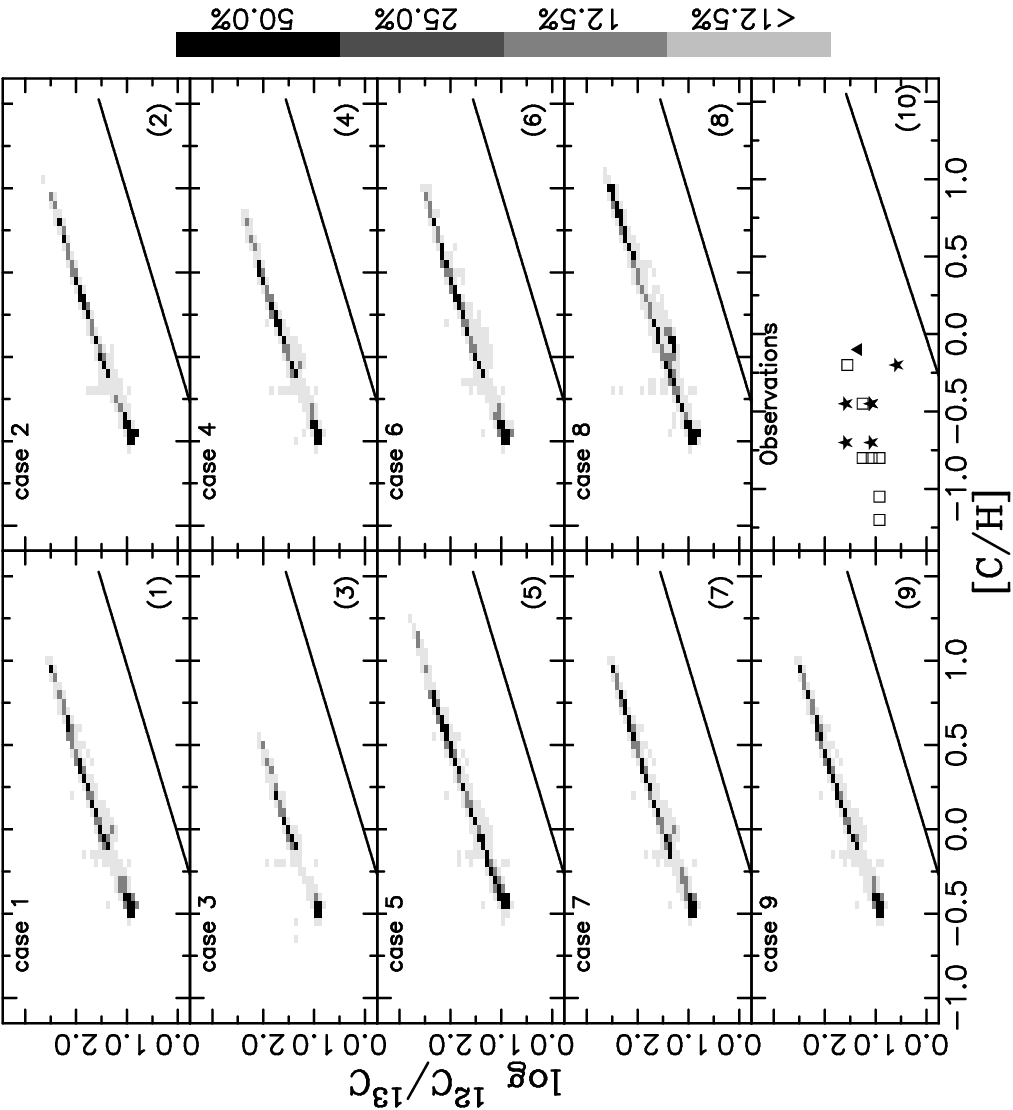}\\
\end{tabular}
\caption{Gray-scale maps of the chemical abundance ratios of
            $\log$ $^{12}$C/$^{13}$C versus [C/H]. The gradations of gray-scale
            correspond to the regions where the number density of systems is,
            respectively,  within 1 -- 1/2,
            1/2 -- 1/4, 1/4 -- 1/8, 1/8 -- 0 of the maximum of
             ${{{\partial^2{N}}\over{\partial {\log {\rm ^{12}C/^{13}C}}}{\partial {\log
            {\rm [C/H]}}}}}$, and blank regions do not contain any stars.
            Solid lines mean $\log$ $^{12}$C/$^{13}$C=$\log$ [C/H]. The left panel of Fig.
            \ref{fig:cc} shows the chemical abundance ratios of the cool giants
            in SSs.  The right panel of Fig. \ref{fig:cc} gives those of the cool giants
            in SSs, where $^{12}$C abundance of all giants with initial masses
            $\leq$ 2$M_\odot$ is artificially decreased by a factor
            of 2.5 after the first dredge-up.
            In the panel of observations, the filled big-stars, the empty squares and triangle represent
            the observational values in \cite{s92}, those in \cite{sm03} and those in \cite{s06}, respectively.
            }
\label{fig:cc}
\end{figure*}

Using the infrared spectra of SSs, \cite{s92} and \cite{sm03} gave
the C abundance and $^{12}$C/$^{13}$C isotopic ratio of the cool
giants in 13 SSs with $\pm$0.3 errors. Using high-resolution
near-infrared spectra and the method of the standard local thermal
equilibrium analysis and atmospheric models, \cite{s06} calculated
the abundance of the symbiotic star CH Cyg with the errors less than
0.3 dex for all the elements. Their observational data are shown in
panels (10) of Fig. \ref{fig:cc}, where [C/H] means the relative
abundances to the solar, that is, $[{\rm C/H}]={\rm \log C/H-\log
C_\odot/H_\odot}$. The left panel of Fig. \ref{fig:cc} shows that
$^{12}$C/$^{13}$C and [C/H] of the cool giants in SSs are much
higher than those from the observations. \cite{s92} used the
synthetic spectra calculated for M giants by \cite{l91} in which
they analyzed M giants and obtained a low mean [C/H]=$-0.64\pm0.29$.
\cite{l91} suggested that the standard mixing named as the first
dredge-up (See \citealt{ir83}) is insufficient to explain
atmospheric abundances in M giants. In our work, we use the standard
mixing which only includes the first dredge-up. $^{12}$C abundance
on stellar surface is reduced by approximately 30\% after the first
dredge-up during the FGB \citep{ir83}.

\cite{gs00} determined Li, C, N, O, Na and Fe abundances and
$^{12}$C/$^{13}$C isotopic ratios of 62 metal-poor field giants.
They suggested that there are two distinct mixing stages along the
red giant branch evolution of small mass field stars: (i) The first
dredge-up, that is, the standard mixing; (ii) The second mixing
episode which is also called the thermohaline mixing in \cite{cz07}
when the giants have a brighter luminosity. The thermohaline mixing
can decrease $^{12}$C abundances. \citet{cz07} calculated several
evolution models of a 0.9$M_\odot$ star and obtained the abundances
which are consistent with the observational data in \cite{gs00}.
Recently, \cite{e06,e07} confirmed a possible mechanism for such
non-canonical mixing. Using 3D-modeling of a low-mass star at the
tip of red giant branch, they found that the molecular inversion can
lead to an efficient mixing and Pop I stars between 0.8 and 2.0
$M_\odot$ develop $^{12}$C/$^{13}$C ratios of 14.5$\pm$1.5. The
average of [C/H] in \cite{s92} is about -0.54 and the average
$^{12}$C/$^{13}$C is about 16. It range in our simulations is
between about -0.05 ( case 3) and 0.15 (case 4) for [C/H] and
between about 23.4 (case 3) and 37.2 (case 5) for $^{12}$C/$^{13}$C,
respectively. The main reason of the above disagreement is that the
thermohaline mixing is not included in our simulation. In order to
show the effects of the thermohaline mixing, we artificially
decrease $^{12}$C abundance of all giants with initial masses $\leq$
2$M_\odot$ by a factor of 2.5 after the first dredge-up. The results
are given in the right panel of Fig. \ref{fig:cc}. The averages of
[C/H] are between -0.31 (case 3) and -0.102 (case 5) and the
averages of $^{12}$C/$^{13}$C are between 12.3 (case 3) and 19.5
(case 5). Our results are then close to the observations of
\cite{s92}.

We believe that the thermohaline mixing can occur and give
significantly different results although the observational samples
of $^{12}$C/$^{13}$C vs. [C/H] of the cool giants in SSs is small.
However, due to poor knowledge on the thermohaline mixing, we can
not carry out a detailed simulation. In the following sections, we
do not discuss its effects any more.
\subsection{Chemical Abundance of Symbiotic Nebulae}
In this subsection, we discuss the chemical abundances of the
symbiotic novae. In order to clarify the nature of the symbiotic
nebulae, we need the observational abundances of novae and planetary
nebulae. In this paper, the nova abundances come from \cite{lt94} in
which they analyzed the abundances of 18 classical novae with a
factor of $\sim$ 2-4 uncertainty. The planetary nebulae abundances
are obtained from \cite{pb06} in which they gave the chemical
abundance of 26 planetary nebulae within a 30\% uncertainty in the
abundance determination.

\subsubsection{\rm C/N vs. O/N}
\begin{figure*}
\begin{tabular}{c@{\hspace{3pc}}c}
\includegraphics[totalheight=3.2in,width=4in,angle=-90]{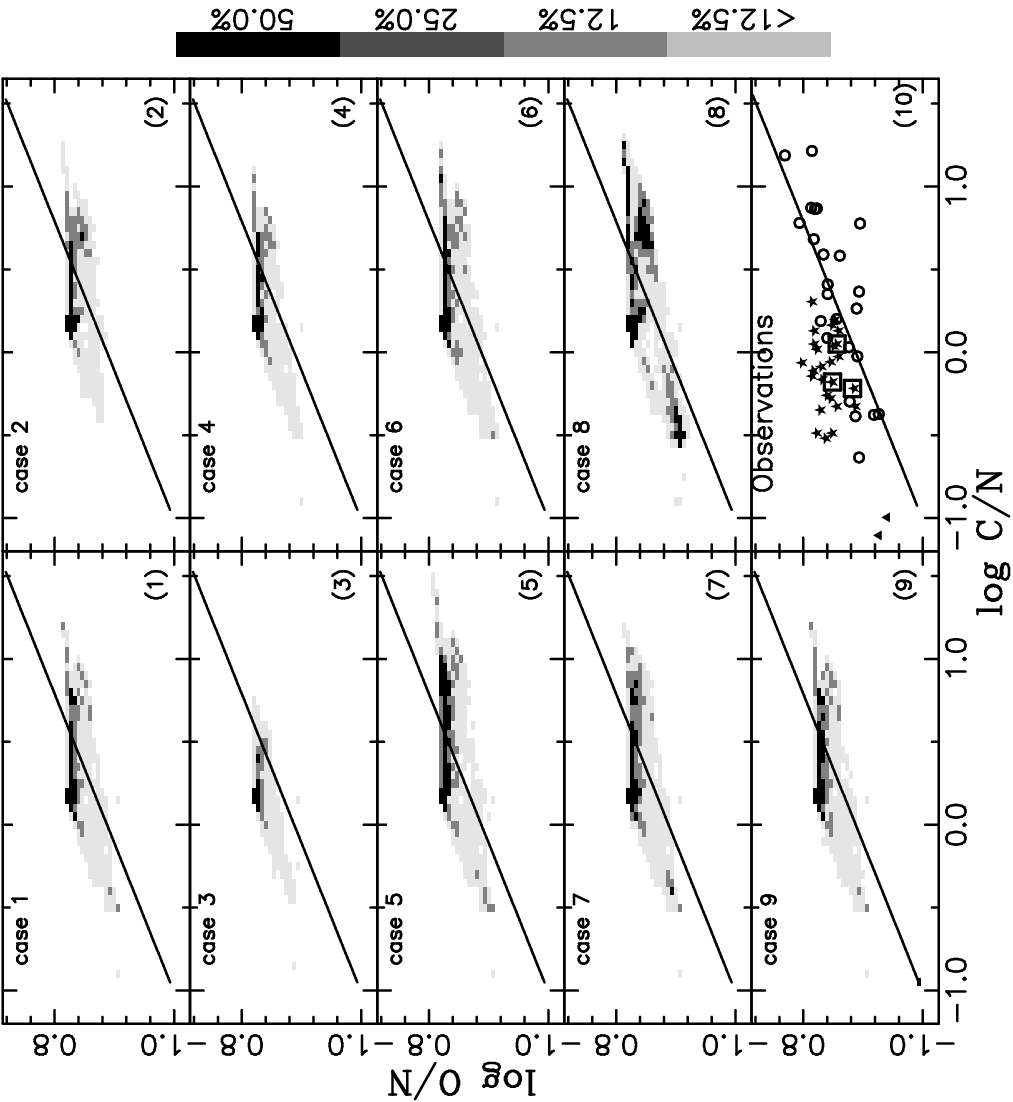}&
\includegraphics[totalheight=3.2in,width=4in,angle=-90]{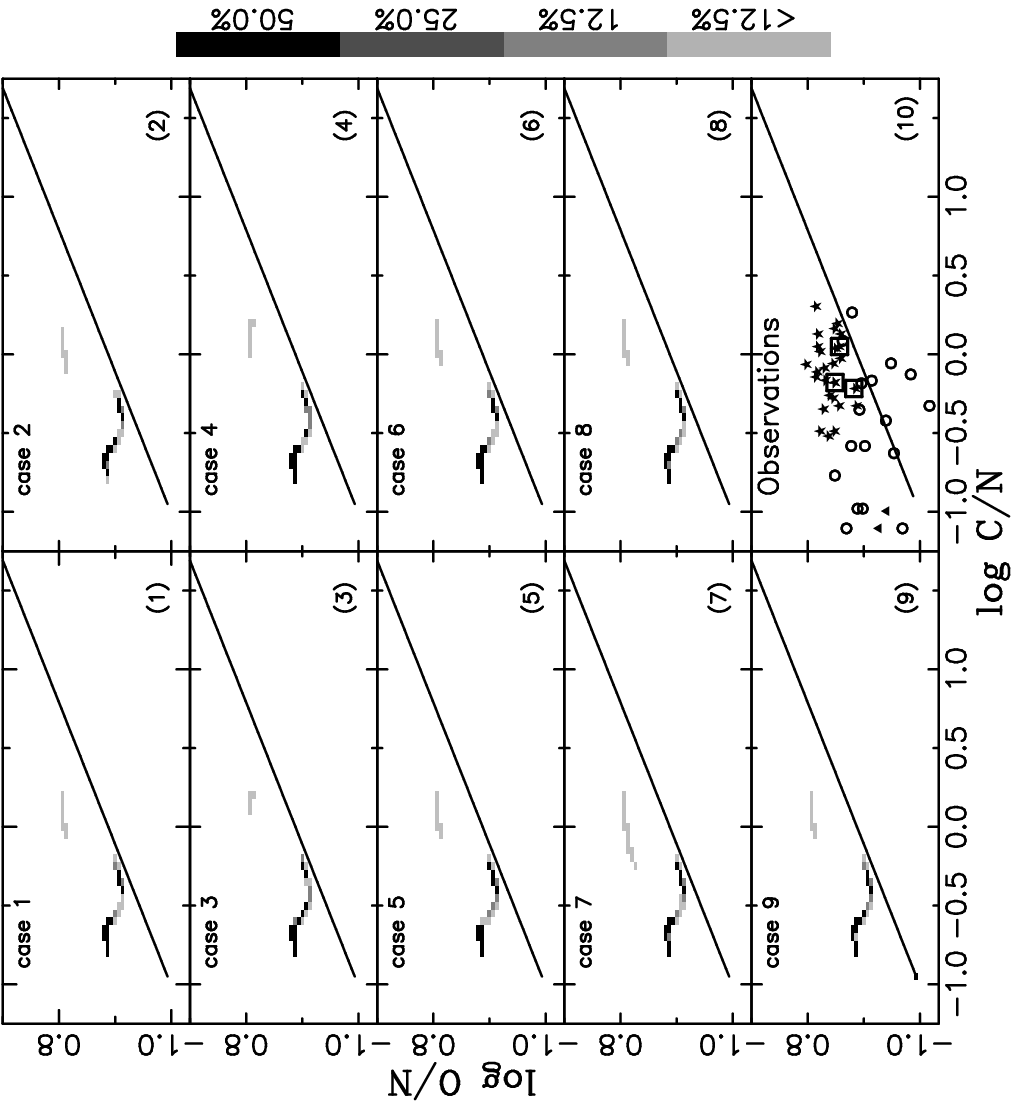}\\
\end{tabular}
\caption{Gray-scale maps of the chemical abundance ratios of
            $\log$ C/N versus $\log$ O/N. The gradations of gray-scale
            correspond to the regions where the number density of systems is,
            respectively,  within 1 -- 1/2,
            1/2 -- 1/4, 1/4 -- 1/8, 1/8 -- 0 of the maximum of
             ${{{\partial^2{N}}\over{\partial {\log {\rm C/N}}}{\partial {\log
            {\rm O/N}}}}}$, and blank regions do not contain any stars.
            Solid lines mean $\log$ C/N=$\log$ O/N. The left and right panels of Fig.
            \ref{fig:co} show the chemical abundances of the giants during the symbiotic stage
            and the ejected materials from the hot companions during the symbiotic novae, respectively.
            In the panels of observations, the filled big-stars, the
            filled triangles and the squares represent the observational values of the symbiotic nebulae in \citet{n88},
            those of PU Vul in \citet{vn92} (including C$_{\rm IV}$ and N$_{\rm IV}$ or excluding
            them) and those of Cyg, HM Sge and HBV 475 in
            \cite{ss90}, respectively. The circles in the left and right panels
            (10) of Fig. \ref{fig:co} represent the observational values
            of the planetary nebulae in \citet{pb06} and the novae in \citet{lt94}, respectively.  }
\label{fig:co}
\end{figure*}

Based on UV data, \cite{n88} presented C/N and O/N abundance ratios
of 24 symbiotic nebulae. The errors in the logarithmic (basis of 10)
abundance ratios remain within 0.18. In their sample, there are at
least three symbiotic novae V1016 Cyg, HM Sge and HBV 475. V1016 Cyg
and HBV 475 had their outbursts around 1965 while HM Sge brightened
by 5 mag in 1975. \cite{ss90} calculated their abundance ratios
within an error of 30\%. Their C/N and O/N abundance ratios are
different from those in \cite{n88} within a factor of 30\%.
According to $IUE$ observations, HM Sge is still evolving towards
higher excitation whereas V1016 Cyg and HBV 475 seem to have reached
their maximum excitation \citep{nv90}. The outburst of PU Vul began
late in 1977. \cite{vn92} gave the C/N and O/N abundances ratios in
the early nebular phase. The observational results and our simulated
results are shown in Fig. \ref{fig:co}.

The left and right panels (1)--- (9) of Fig. \ref{fig:co} give the
distribution of the relative CNO abundances of the cool companions
and the ejected materials from the hot components during symbiotic
novae, respectively. Panels (10) in Fig. \ref{fig:co} show the
observational ratios of C/N vs. O/N of symbiotic nebulae, novae and
planetary nebulae.

The left panel of Fig. \ref{fig:co} shows that we can cut the
abundance ratios of C/N vs. O/N of the cool companions in SSs into
three regions. i) The region of C/N $>$ O/N represents that the cool
companions have undergone the deep TDU and have not undergone the
hot bottom burning. The initial masses of the cool companions should
be between about 2.0 or 1.5(case 5) $M_\odot$ and 4.0 $M_\odot$.
Their abundances are similar to those of C-stars and planetary
nebulae; ii) The top region of C/N $<$ O/N represents that the cool
companions have undergone inefficient TDU or have not undergone TDU.
These stars have initial masses lower than 2.0 or 1.5 (case 5)
$M_\odot$. Their abundances are similar to those of M-stars; iii)
The bottom region of C/N $<$ O/N represents that the cool companions
undergo TDU and the hot bottom burning which turns $^{12}$C to
$^{14}$N. Their initial masses are higher than 4.0$M_\odot$.

In the right panel of Fig. \ref{fig:co}, the distribution of C/N vs.
O/N of the ejected materials from the hot companions during
symbiotic novae is composed of two regions. The top region
represents symbiotic novae with accreting ONe WDs. The bottom region
represents those with accreting CO WDs. Compared with the
observations, our results agree with those in \cite{lt94}.

From the left panel of Fig. \ref{fig:co}, we find that the relative
CNO abundances of the cool companions in our simulations are close
to those of the observational symbiotic nebulae. However, comparing
more carefully, one can find that many observed SSs lie in the
transition from normal M giants with initial mass lower than
4$M_\odot$ (no hot bottom burning) to the super giants with higher
initial mass. This agrees with \cite{n88}. If we compare Fig.
\ref{fig:co} with the observations, we find that the abundance
ratios of most of observational symbiotic nebulae are between those
of the cool giants and the ejected materials from the hot companions
during symbiotic novae. This suggests that the compositions of the
symbiotic nebulae may be modified by the ejected materials from the
hot companions.

As the panels (10) of Fig. \ref{fig:co} show, the abundance ratios
of the symbiotic novae Pul Vul and HM Sge are like those of the
novae while the ratios of V1016 Cyg and HBV 475 are similar to those
of the giants. Pul Vul and HM Sge are still staying on the early
phase of the nova outbursts whereas V1016 Cyg and HBV 475 may be
staying on the decline phase. \cite{vn92} predicted that the
chemical abundances of Pul Vul would become closer to those of red
giants in the future. Therefore, we believe that the compositions of
the symbiotic nebulae have been modified by the ejected materials
from the hot companions during symbiotic novae. Can the modification
only last $t_{\rm on}$ like our assumption in \S \ref{sec:syne}?
According to Table \ref{tab:result}, most of SSs are in the cooling
phase of the symbiotic novae and the stable hydrogen burning phase.
The symbiotic novae are only 18.3\% (case 7) and 10\% (case 9) of
total SSs. However, most of the observed symbiotic nebulae have
mixing CNO abundances, which means that the above modification
should be common in all SSs. We showed the C/N and O/N abundance
ratios of several typical SSs which stay in different phases in
order to support the above the view of point.


{\bf Declining phase:} AG Pegasi is currently in decline from a
symbiotic nova outburst which began in about 1850. Based on
multi-shell radio emission from AG Pegasi in \cite{k91}, the nebula
can be divided into three parts. The outer nebula is thought to be
the remnant of the 1850 outburst. The intermediate nebula may
represent the pre-eruption mass loss from the cool component. Using
a new colliding winds models including the effects of orbital
motion, \cite{kt07} suggested that the inner nebula originates from
the colliding winds and has an expansion velocity of about 50 km
s$^{-1}$. It takes about 1000 yr for the inner nebula to reach the
region of the outer nebula. In general, the timescale of the cooling
phases $t_{\rm cool}$ is several hundred years. Therefore, during
the cooling phase after the thermonuclear runaways, the ejected
materials from the hot components have an effect on the chemical
compositions of the symbiotic nebulae.

{\bf Multiple outbursts:} \cite{so06} discussed the outburst of
symbiotic system Z And occurring in 2000-2002. They believed that
the outburst in 2000-2002 results from the enhanced shell burning
triggered by a disk instability. During the burst occurring in
2000-2002, a shell of material blown from the surface of the WD is
found. The blown material may result from an optically thick wind.
Both Z And and CH Cyg are SSs which have undergone multiple
outbursts. Since 1964, CH Cyg has displayed a number of outbursts.
\cite{s06} showed both the photospheric features of the cool giant
and the nebular emission lines in CH Cyg. The average C/N and O/N
ratios of the cool giants in CH Cyg are respectively 1.6 and 4.0,
which are typical abundance ratios of M giants. The average ratios
of the symbiotic nebulae in CH Cyg are respectively 0.57 and 2.0,
which are similar to those of super giant stars and are between
those of giants and novae. The significant discrepancy of C/N and
O/N ratios derived from these two components suggest that the
chemical abundances of the nebulae in CH Cyg are modified by the
ejected materials from the hot components.

{\bf Quiescent phase:} Sy Mus lacks any outburst activity and it is
in the stable hydrogen burning phase \citep{m07}. \cite{n88} showed
that the C/N and O/N ratios in its nebulae are 0.97 and 1.9,
respectively. These ratios are similar to those of super giant stars
and are between those of giants and novae. This means that the
chemical abundances of the nebulae of Sy Mus in the quiescent phase
are modified by the ejected materials from the hot components.

Therefore, we suggest that it is quite common in SSs that the
chemical abundances of the symbiotic nebulae are modified by the
ejected materials from the hot WD surfaces.

\subsubsection{\rm Ne/O vs. N/O}
\begin{figure*}
\begin{tabular}{c@{\hspace{3pc}}c}
\includegraphics[totalheight=3.2in,width=4in,angle=-90]{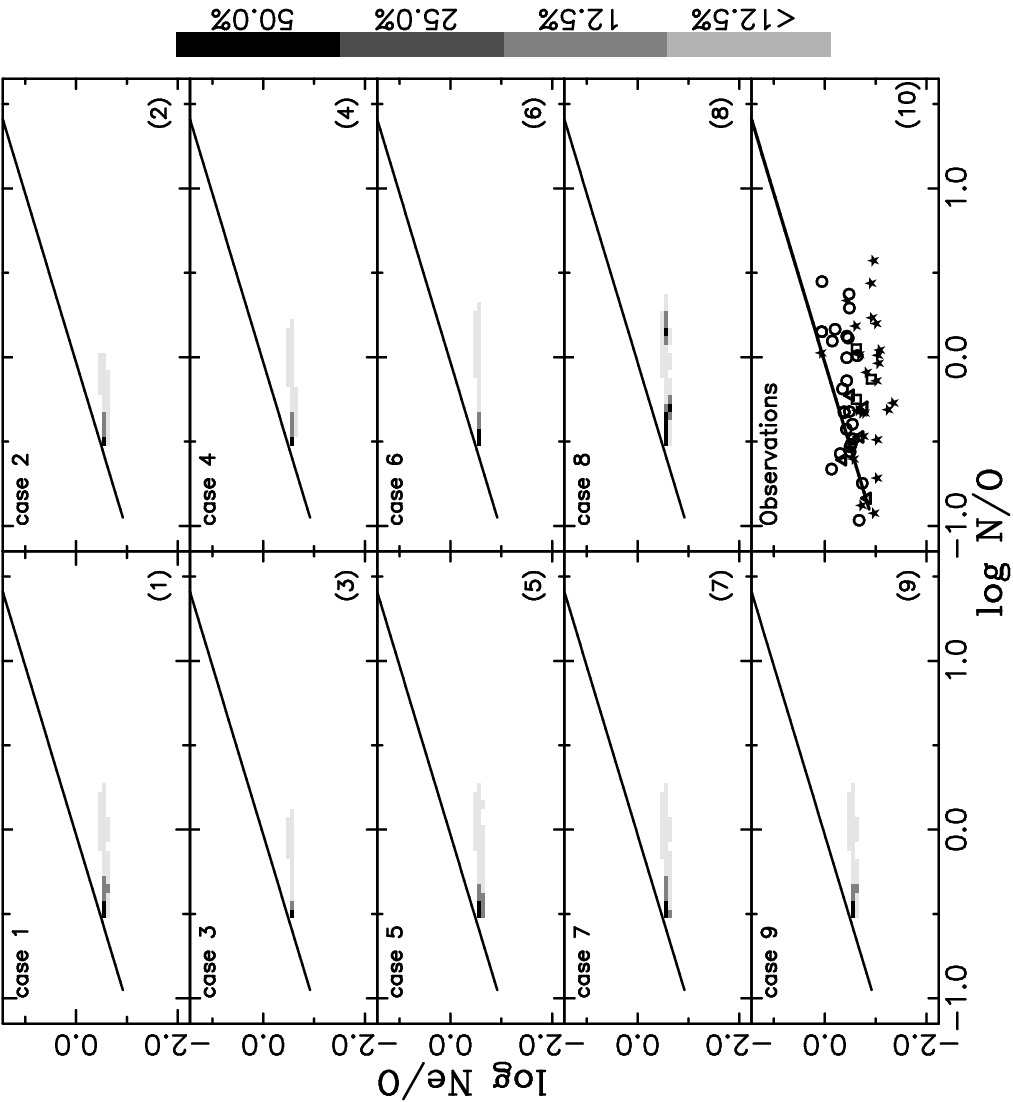}&
\includegraphics[totalheight=3.2in,width=4in,angle=-90]{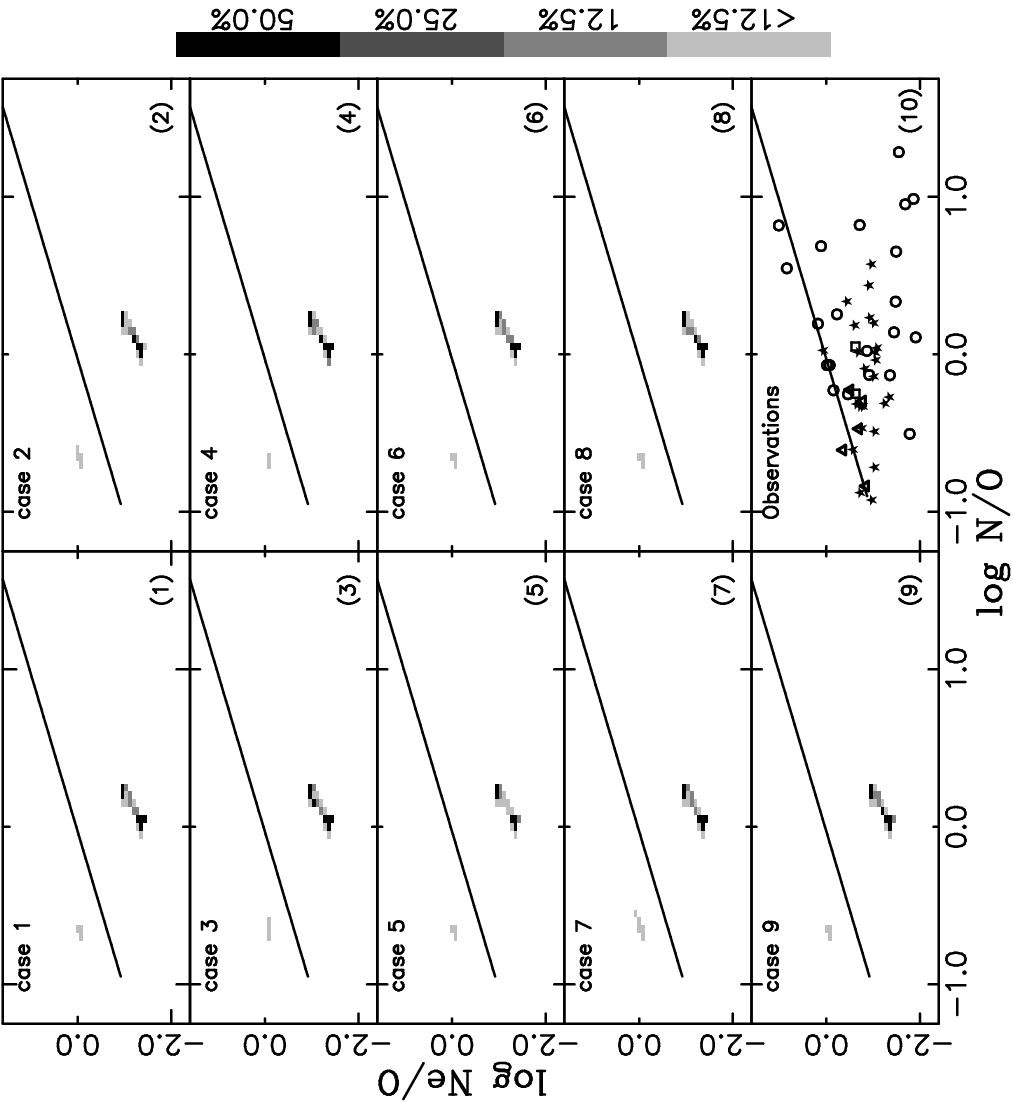}\\
\end{tabular}
\caption{Gray-scale maps of the chemical abundance ratios of
            $\log$ Ne/O versus $\log$ N/O.
            Solid lines mean $\log$ Ne/O=$\log$ N/O. The left and right panels of Fig.
            \ref{fig:neo} show the chemical abundances of the giants during the symbiotic stage
            and the ejected materials from the hot companions during the symbiotic novae, respectively.
            In the panels of the observations, the filled big-stars, the
            filled triangles and the squares represent the observational values of the symbiotic nebulae in \citet{lc05},
            those of 5 symbiotic nebulae in \cite{dc92} and those of V1016 Cyg and HM Sge in
            \cite{ss90}, respectively. The circles in the left and right panels
            (10) of Fig. \ref{fig:neo} represent the observational values
            of the planetary nebulae in \citet{pb06} and the novae in \citet{lt94}, respectively.
            }
\label{fig:neo}
\end{figure*}

According to optical data, \cite{lc05} calculated He abundance and
the relative abundances N/O, Ne/O and Ar/O of 43 symbiotic nebulae.
We first discuss Ne/O vs. N/O . The mean errors in the logarithmic
(basis of 10) abundance ratios are $\sim \pm$ 0.1. \cite{dc92}
showed the ratios of N/O and Ne/O in 5 symbiotic nebulae within an
error of 0.15. Using the observational data, we note that the
observational results for same SSs in different literature are quite
different. The ratios of N/O and Ne/O of V1016 Cyg in \cite{dc92}
differ by a factor of 4 from those in \cite{ss90}.

Fig. \ref{fig:neo} gives the distributions of the abundance ratios
of Ne/O vs. N/O in all cases and observations. The left and right
panels of Fig. \ref{fig:neo} show Ne/O vs. N/O of the cool giants in
SSs and the ejected materials from the hot components during
symbiotic novae, respectively. The right panel of Fig. \ref{fig:neo}
shows that our results of the ejected materials from the hot
companions during the symbiotic novae are less scattered than the
observations of the novae in \cite{lt94}. There are mainly two
reasons. One reason is too small numerical samples in \cite{kp97}
and \cite{jh98}. Another reason is a high uncertainty in
observations with a factor of 2-4. In the right panel of Fig.
\ref{fig:neo}, the distributions of Ne/O and N/O are obviously cut
into two regions. The left-up region of higher Ne/O originates from
the ejected materials from the hot companions during the symbiotic
novae with accreting ONe WDs and the middle-down region of lower
Ne/O corresponds to the ejected materials from the hot companions
during the symbiotic novae with accreting CO WDs.

Panels (10) of Fig. \ref{fig:neo} show that the abundance ratios of
Ne/O vs. N/O in symbiotic nebulae are basically between those in the
planetary nebulae and the novae. Our simulations also support this
result. Therefore, the distributions of Ne/O vs. N/O reconfirm that
the symbiotic nebulae are modified by the ejected materials from the
hot components.

\subsubsection{\rm He/H vs. N/O}
\begin{figure*}
\begin{tabular}{c@{\hspace{3pc}}c}
\includegraphics[totalheight=3.2in,width=4in,angle=-90]{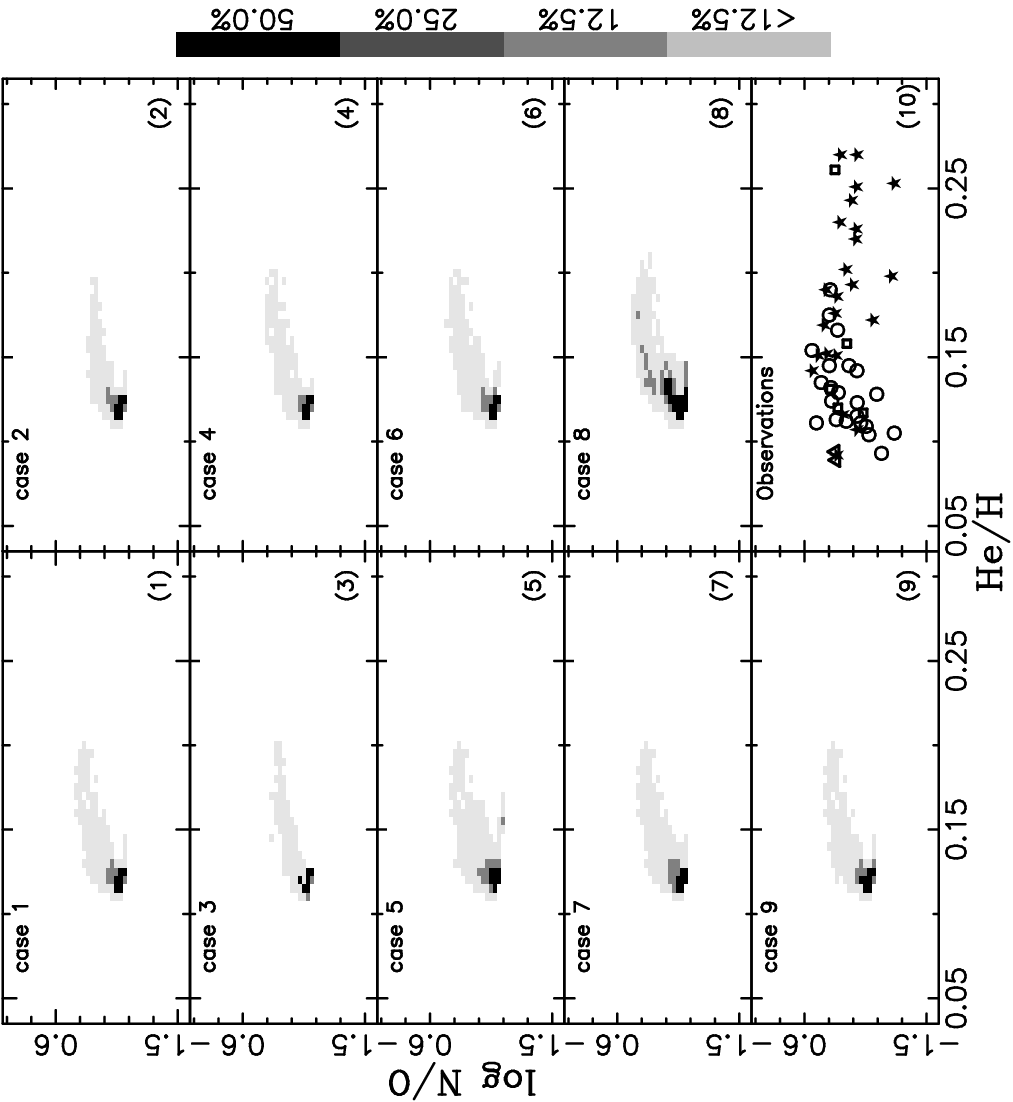}&
\includegraphics[totalheight=3.2in,width=4in,angle=-90]{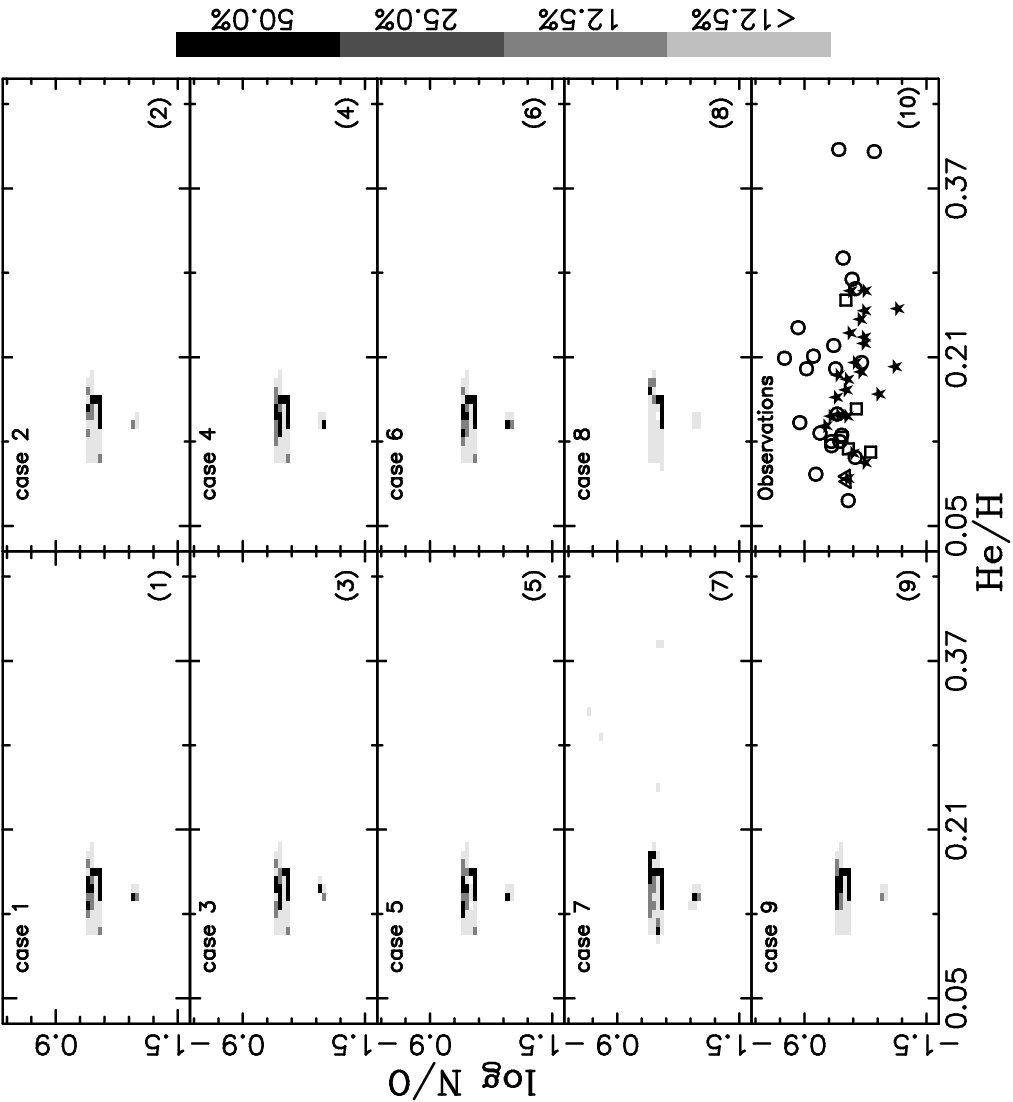}\\
\end{tabular}
\caption{Gray-scale maps of the chemical abundance ratios of He/H
            versus $\log$ N/O.
            The left and right panels of Fig. \ref{fig:he} show the chemical
            abundances of the giants during symbiotic stage
            and the ejected materials from the hot companions during the symbiotic novae, respectively.
            In the panels of the observations, the filled big-stars, the squares and the
            triangles represent the observational values of the symbiotic nebulae in \citet{lc05},
            those of 5 symbiotic nebulae in \cite{cd94} and those of V1016 Cyg and HM Sge in
            \cite{ss90}, respectively. The circles in the left and right panels
            (10) of Fig. \ref{fig:he} represent the observational values
            of the planetary nebulae in \citet{pb06} and the novae in \citet{lt94}, respectively.
            }
\label{fig:he}
\end{figure*}

\cite{lc05} showed the distributions of the observational $\log$ O/N
vs. He/H for 43 symbiotic nebulae. The mean error of $\log$ O/N is
about 0.1 and the mean error of He/H is within $\sim$ 0.03.
\cite{cd94} gave those of 5 symbiotic nebulae within an error of
0.038.

As the panels (10) of Fig. \ref{fig:he} show, there are many
symbiotic nebulae whose helium abundances are much higher than those
of the observational planetary nebulae in \cite{pb06}. The
enhancement of the helium abundance in the stellar surface mainly
depends on the third dredge-up. However, the left panels (1)--(9) of
Fig. \ref{fig:he} show that the He/H ratios in the stellar surface
of the cool giants in all SSs are basically lower than 0.2 in our
simulations. The enhancement of the helium abundance in the
symbiotic nebulae should originate from the ejected materials during
the symbiotic novae. The observational He/H ratios of the classical
novae in \cite{lt94} range from 0.07 to 0.40. The observational
distribution of He/H vs. $\log$ N/O also support that the symbiotic
nebulae are modified by the ejected materials from the hot
components. Unfortunately, as the right panels (1)--(9) of Fig.
\ref{fig:he} show, our simulations do not offer high He/H ratios for
the ejected materials from the hot components. A possible
explanation is: if the mass-accretion rates of WD accretors are
higher than the critical value $\dot{M}_{\rm WD}$ (See Eq. (24) in
Paper I), the nova outbursts are weak and a helium layer may be left
on the surface of WD accretors. We call these novae as weak
symbiotic novae (Paper I). If the mass-accretion rates of WD
accretors are lower than $\dot{M}_{\rm WD}$, the nova outbursts are
so strong that the WDs are eroded. Therefore, no helium layer is
left. These novae are called as strong symbiotic novae. A typical
symbiotic star can undergo dozens of nova outbursts (Paper I). For
the weak symbiotic novae, a helium layer may be left on the surface
of WD accretors after each outburst. When the next outburst occurs,
the remnant helium can be dredged up and ejected \citep{i92}. In
addition, most WDs have been traditionally found to belong to one of
the following categories: those with a hydrogen-rich atmosphere (the
DAs) and those with a helium-rich atmosphere (the DBs). DA white
dwarfs constitute about the 80\% of all observed WDs. They have a
thin hydrogen layers whose mass is lower than about $10^{-4}M_{\rm
WD}$. Also, there is a helium layer (their mass is about
$10^{-2}M_{\rm WD}$) under the hydrogen layer \citep{ab98}.
\cite{it85} calculated the distributions of helium, carbon and
oxygen through the surfaces layers of a 1.05 $M_\odot$ DB WD coming
from a case B mass-transfer event. They found the amount of helium
near the surface is only about $10^{-3}M_\odot$. \cite{pm07} tested
the evolution of the DB white dwarf GD 358. They suggested that
binary evolution describes better GD 358 and obtained the helium
layer with a mass $10^{-5.66}M_{\rm WD}$. Therefore, when the first
nova outbursts occurs, the hydrogen shell flash will penetrate
inward to the region where the abundance of H is $\sim$ 0.01 and a
certain amount of the helium is dredged up \citep{i92}. In the above
situations, helium is overabundant. Our calculations may
underestimate helium abundances of the ejected materials from the
hot components during the symbiotic novae.

The helium of the symbiotic novae V1016 Cyg and HM Sge in
\cite{ss90} is not overabundant. They may have undergone several
strong symbiotic novae so that the WD accretors have no the helium
layer. Their helium should be similar to our simulated result.

We suggest that the helium enhancement of the symbiotic nebulae
stems from the modification of the ejected materials from the WDs
with a helium layer. In our simulations, the ratios of the SSs being
in the weak symbiotic novae and their decline phases to total SSs
are between 25\% (case 2) and 38\% (case 9). The helium element of
these symbiotic nebulae should be overabundant. The ratio of the
symbiotic nebulae whose He/H is higher than 0.2 to the total of the
symbiotic nebulae in \cite{lc05} is 33.3\%, which are in agreement
with our results.

\section{Conclusion}
We have performed a detailed study of the chemical abundances in SSs
and drew several important conclusions.

1. The initial-final mass relation has great effects on SSs'
population and especially on the symbiotic novae with massive WD
accretors. A steep initial-final mass relation result in an
overestimated occurrence rate of symbiotic novae.

2. The number ratios of O-rich SSs to C-rich SSs in our simulations
are between 3.4 and 24.1, and they are sensitive to TDU efficiency
$\lambda$ and the terminal velocity of stellar wind $v(\infty)$. Our
simulation may have underestimated the terminal velocity of stellar
wind $v(\infty)$ of C-rich giants and the mass-loss rate of the cool
giants in SSs. The number ratio of SSs with $extrinsic$ C-rich cool
giants to all SSs with C-rich cool giants is between 2.1\% and
22.7\%. The TDU efficiency $\lambda$, the common envelope algorithm
and the mass-loss rate have a great effect on it.

3. Comparing our $^{12}$C/$^{13}$C vs. [C/H] of the cool giants in
SSs with those of the observations, we infer that thermohaline
mixing in low-mass stars should exist. Its effect on the chemical
abundances is very significant.

4. The distributions of O/N vs. C/N,  Ne/O vs. N/O and He/H vs. N/O
of the symbiotic nebulae indicate that it is quite common in SSs
that the nebular chemical abundances are modified by the ejected
materials from the hot components.

5. Helium abundances in the symbiotic nebulae during the symbiotic
novae are determined by whether the WD accretors have a helium layer
or not in their surfaces. If they have a helium layer, helium is
overabundant. If the WD accretors have only undergone strong
symbiotic nova outbursts, they have no helium layer. The helium in
the symbiotic nebulae ( like V1016 Cyg and HM Sge ) are not
overabundant.

\section*{Acknowledgments}
We thank an anonymous referee for his/her comments which helped to
improve the paper. We are grateful to Dr. Izzard for providing his
Doctor's thesis and J. Miko{\l}ajewska for providing us a
compilation of orbital periods of SSs. LGL thanks Ms. Miranda
Beckham for correcting English language of the manuscript. This work
was supported by the Chinese National Science Foundation under
Grants Nos. 10647003, 10521001 and 10763001, the Doctor Foundation
of Xinjiang University (BS060109) and the Foundation of Xinjiang
University (070195).
\bibliographystyle{mn2e}
\bibliography{zhongmuads}

\appendix
\label{sec:appa}
\section[]{Synthesis TP-AGB Evolution} Stellar evolution from the zero
age main sequence up to the first thermal pulse is dealt with in the
rapid evolution code of \citet{h00}. The changes of the chemical
abundances on the stellar surface during the giant branch phase (the
first dredge up) and early asymptotic giant branch phase (the second
dredge up) can be represented by simple fitting formulae in
\cite{i04} and \cite{I04}. After the first thermal pulse, we use a
synthetic model for TP-AGB.

\subsection{The initial abundances} The
initial abundances ({\it i.e.} of zero age main sequence stars) are
taken from \citet{AG89} for $Z=0.02$. The
following shows the initial abundances by mass fractions:\\
$^1$H=0.68720; \  \ \ \ \  \ \ \ \ \ \ \ \ \ \ \ \ \ $^4$He=0.29280 ;\\
$^{12}$C=2.92293$\times10^{-3}$; \  \ \ \ \  \ \ \ $^{13}$C=4.10800$\times10^{-5}$;\\
$^{14}$N=8.97864$\times10^{-4}$; \  \ \ \  \ \ \ \
$^{15}$N=4.14000$\times10^{-6}$;\\
$^{16}$O=8.15085$\times10^{-3}$; \  \ \ \  \ \ \ \  $^{17}$O=3.87600$\times10^{-6}$;\\
$^{20}$Ne=2.29390$\times10^{-3}$; \  \ \ \  \ \ \ $^{22}$Ne=1.45200$\times10^{-4}$.\\
\subsection{Core mass at the first thermal pulse} Using the rapid evolution code of
\citet{h00}, we can obtain the stellar mass ($M_{\rm 1TP}$) at the
first thermal pulse. The core mass at the first thermal pulse,
$M_{\rm c, 1TP}$ is taken from \cite{k02}:
\begin{equation}
M_{\rm c, 1TP}=[-p_1(M_0-p_2)+p_3]f+(p_4M_0+p_5)(1-f),
\label{eq:mc1tp}
\end{equation}
where $f=(1+e^{(\frac{M_0-p_6}{p_7})})^{-1}$, $M_0$ is the initial
mass in solar unit and the coefficients  $p_1$, $p_2$, $p_3$, $p_4$,
$p_5$, $p_6$ and $p_7$ are in Table 6 of \cite{k02}.
\subsection{Luminosity, radius, interpulse period, and evolution of core mass}
We use the prescriptions of \cite{i04}. The luminosity is taken as
the value calculated from Eq.(29) in \cite{i04}. We define the
radius $R$ as $L=4\pi \sigma R^2T^4_{\rm eff}$, where $\sigma$ is
the Stefan-Boltzmann constant and $T_{\rm eff}$ is the effective
temperature of the star. The radius is taken the value calculated
from Eq.(35) in \cite{i04}. The interpulse period $\tau_{\rm ip}$
is:
\begin{equation}
{\rm log}_{10}(\tau_{\rm ip}/\rm yr)=a_{28}(M_{\rm
c}/M_\odot-b_{28})-10^{c_{28}}-10^{d_{28}}+0.15\lambda^2,
\label{eq:tip}
\end{equation}
where the third dredged-up (TDU) efficiency $\lambda$ is defined in
\S \ref{sec:lamb}, and
the coefficients are:\\
$a_{28}=-3.821,$\\
$b_{28}=1.8926,$\\
$c_{28}=-2.080-0.353Z+0.200(M_{\rm env}/M_\odot+\alpha-1.5),$\\
$d_{28}=-0.626-70.30(M_{\rm c,1TP}/M_\odot-\zeta)(\Delta M_{\rm
c}/M_\odot),$\\where $M_{\rm env}$ represents the envelope mass,
$\alpha$ is the mixing length parameter and equals to 1.75,
$\zeta={\rm log}(Z/0.02)$, and $\Delta M_{\rm c}$ is the change in
core mass defined as $\Delta M_{\rm c}=M_{\rm c}-M_{\rm c, 1TP}$.

\subsection{The minimum core mass for TDU and the TDU efficiency}
\label{sec:lamb} TDU can occur only  for stars above a certain core
mass $M^{\rm min}_{\rm c}$.  \cite{gj93} took $M^{\rm min}_{\rm c}$
as a constant 0.58. \cite{k02} found that $M^{\rm min}_{\rm c}$
depends on stellar mass and metallicity and gave a fitting formula
by
\begin{equation}
M^{\rm min}_{\rm c}=a_1+a_2M_0+a_3M_0^2+a_4M_0^3 \label{eq:mcmin},
\end{equation}
where coefficients $a_1$, $a_2$, $a_3$ and $a_4$ are shown in Table
7 of \cite{k02} and $M_0$ is the initial mass in solar unit.
According to the observed carbon luminosity function in the
Magellanic clouds, \cite{mg07} considered that the $M^{\rm min}_{\rm
c}$ predicted  by \cite{k02} is high. In this work, we carry out
various numerical simulations (See Table
\ref{tab:case}).\\
(i)Like \cite{gj93}, $M^{\rm min}_{\rm c}$=0.58$M_\odot$;\\
(ii)Following \cite{k02}, we take Eq. (\ref{eq:mcmin}) as $M^{\rm min}_{\rm c}$;\\
It should be recalled that $M^{\rm min}_{\rm c}$=$M_{\rm c, 1TP}$ if
 stellar initial mass $M_{\rm initial}\geq 4M_\odot$ \citep{k02,i04}
or $M^{\rm min}_{\rm c}<M_{\rm c, 1TP}$.

The TDU efficiency is defined by $\lambda=\frac{\Delta M_{\rm
dred}}{\Delta M_{\rm c}}$, where $ \Delta M_{\rm dred}$ is the mass
brought up to the stellar surface during a thermal pulse.  $\lambda$
is a very uncertain parameter. \cite{k02} showed a relation of
$\lambda$:
\begin{equation}
\lambda(N)=\lambda_{\rm max}[1-\exp(-N/N_{\rm r})], \label{eq:lamb}
\end{equation}
where $\lambda$ gradually increases towards an asymptotic
$\lambda_{\rm max}$ with $N$ (the progressive number of thermal
pulsation) increasing. $N_{\rm r}$ is taken from Eq. (49) in
\cite{i04} which reproduces the results for $N_{\rm r}$ in Table 5
of \cite{k02}. $\lambda_{\rm max}$ is given by (See Eq.(6) in
\cite{k02}):
\begin{equation}
\lambda_{\rm max}=\frac{b_1+b_2M_0+b_3M^3_0}{1+b_4M_0^3},
\label{eq:lambmax}
\end{equation}
where coefficients are shown in Table 8 of \cite{k02}. \cite{gj93}
took $\lambda$ as a constant 0.75. In order to test the influences
of $\lambda$ on our results, we carry out numerical simulations with
various $\lambda$ (See Table \ref{tab:case}).\\
(i)Following \cite{k02}, $\lambda$=Eq. (\ref{eq:lamb});\\
(ii)Following \cite{gj93}, $\lambda$=0.75;\\
(iii)Simulating a small TDU efficiency, $\lambda$=0.5.

\subsection{Inter-shell abundances}
During every thermal pulse, the dredged mass $\Delta M_{\rm dred}$
mixes into stellar envelope. Based on the nucleosynthesis
calculations by \citet{bs88}, \cite{mg07} gave the fit of the
abundances of $^4$He,
$^{12}$C and $^{16}$O in the inter-shell region:\\
For $\Delta M_{\rm c}\leq$0.025$M_\odot$:\\
$^{\rm inter}X(^{4}{\rm He})=0.95+400(\Delta M_{\rm c})^2-20.0\Delta M_{\rm c}$\\
$^{\rm inter}X(^{12}{\rm C})=0.03-352(\Delta M_{\rm c})^2+17.6\Delta M_{\rm c}$\\
$^{\rm inter}X(^{16}{\rm O})=-32(\Delta M_{\rm c})^2+1.6\Delta M_{\rm c}$.\\
 For $\Delta M_{\rm c}>$0.025$M_\odot$:\\
$^{\rm inter}X(^{4}{\rm He})=0.70+0.65(\Delta M_{\rm c}-0.025)$\\
$^{\rm inter}X(^{12}{\rm C})=0.25-0.65(\Delta M_{\rm c}-0.025)$\\
$^{\rm inter}X(^{16}{\rm O})=0.02-0.065(\Delta M_{\rm c}-0.025)$.\\

However, there is some debate of the exact composition in the
inter-shell region. \cite{i04} gave some fits of inter-shell
abundance, in which inter-shell elements include $^{4}$He, $^{12}$C,
$^{16}$O and $^{22}$Ne. Details can be seen from \S 3.3.3 in
\cite{i04}. Their models did not obtain high values of inter-shell
$^{16}$O such as 2\% reported by \citet{bs88}. The typical
inter-shell abundances (5 $M_\odot$, $Z=0.02$) are $^{4}$He=0.74,
$^{12}$C=0.23, $^{16}$O=0.005 and $^{22}$Ne=0.02.

In our work, we carry out simulations with various inter-shell
abundances.

\subsection{The hot bottom burning}
\label{sec:hbb}

If the hydrogen envelope of an AGB star is sufficiently massive, the
hydrogen burning shell can extend into the bottom of the convective
region. This process is called as hot bottom burning(HBB). For HBB,
we use a treatment similar to that in \cite{gj93}. For the model of
\citet{ir83}, \cite{gj93} gave the most suitable parameter for the
fraction of newly dredged up matter exposed to the high temperatures
at the bottom of envelope $f_{\rm HBB}=0.94$, the fraction of
envelope matter mixed down to the bottom of the envelope $f_{\rm
bur}=3\times10^{-4}$, and the exposure time of matter in the region
of HBB $t_{\rm HBB}=0.0014\tau_{\rm ip}$. The temperature at the
bottom of convective envelope $T_{\rm bce}$ is given by Eq. (37) of
\cite{i04}.

For the drudged up mass, the amounts of material added to the
envelope are:
\begin{equation}
\begin{array}{ll}
\Delta^4{\rm He}=&^{\rm inter}X_4\Delta M_{\rm dred}\\
\Delta^{12}{\rm C}=&[(1-f_{\rm HBB})^{\rm
inter}X_{12}\\
&+\frac{f_{\rm HBB}}{t_{\rm HBB}}
\int^{t_{\rm HBB}}_{0}X^{\rm HBB}_{12}(t){\rm d}t]\Delta M_{\rm dred}\\
\Delta^{13}{\rm C}=&[\frac{f_{\rm HBB}}{t_{\rm HBB}}
\int^{t_{\rm HBB}}_{0}X^{\rm HBB}_{13}(t){\rm d}t]\Delta M_{\rm dred}\\
\Delta^{14}{\rm N}=&[\frac{f_{\rm HBB}}{t_{\rm HBB}}
\int^{t_{\rm HBB}}_{0}X^{\rm HBB}_{14}(t){\rm d}t]\Delta M_{\rm dred}\\
\Delta^{16}{\rm O}=&[(1-f_{\rm HBB})^{\rm
inter}X_{16}\\
&+\frac{f_{\rm HBB}}{t_{\rm HBB}} \int^{t_{\rm HBB}}_{0}X^{\rm
HBB}_{16}(t){\rm
d}t]\Delta M_{\rm dred} ,\\
\end{array}
\label{eq:hbb}
\end{equation}
where $X^{\rm HBB}(t)$ are calculated by the way of Clayton's CNO
bicycle\citep{c83}. All details of Clayton's CNO bicycle can be seen
from \citet{c83,gj93,i04}. The CNO bicycle can be splited into CN
cycle and ON cycle. The timescales in ON cycle are many thousands of
times longer than those required to bring the CN cycle into
equilibrium. Even in the most massive AGB stars undergoing vigorous
HBB, the ON cycle never approaches equilibrium. Therefore, the
effects of HBB mainly turn $^{12}$C into $^{14}$N and the abundance
of $^{16}$O is not changed much. In calculating Clayton's CNO
bicycle, the density at the base of the convective envelope is given
by Eq. (42) of \cite{i04} and the analytic expressions for the
nuclear reaction rates in Clayton's CNO bicycle are taken from
\citet{cf88}. The initial conditions of $X^{\rm HBB}(t)$ are $X^{\rm
HBB}(t=0)=^{\rm inter}X$ for $^{12}{\rm C}$ and $^{16}{\rm O}$,
while $X^{\rm HBB}(t=0)=0$ for $^{13}{\rm C}$ and $^{14}{\rm N}$.

After every thermal pulse, the chemical abundances of stellar
envelope $X^{\rm new}$ are
\begin{equation}
X^{\rm new}=\frac{X^{\rm old}M_{\rm env}(1-f_{\rm bur})+\Delta
X+\frac{f_{\rm bur}M_{\rm env}}{t_{\rm HBB}}\int_0^{t_{\rm
HBB}}X(t){\rm d}t} {M_{\rm env}+\Delta M_{\rm dred}},
\end{equation}
where $\Delta X$ is given by Eq. (\ref{eq:hbb}) and the initial
conditions of $X(t)$ are $X(t=0){\rm }=X^{\rm old}{\rm }$.

\subsection{Mass loss}
\label{sec:ml} The mass-loss rate of the cool giant during AGB phase
has a great effect on the population of SSs and the chemical
evolution of the stellar surface. We consider two
laws of the mass-loss rate in this work. \\
(i)A mass-loss relation suggested by \citet{vw93} based on the
observations, given as
\begin{equation}
\begin{array}{l}
\log \dot{M}= -11.4+0.0123(P-100\max(M/M_\odot-2.5, 0.0)),
\end{array}
\label{eq:vwml}
\end{equation}
where $P$ is the Mira pulsation period in days given by
\begin{equation}
\begin{array}{l}
\log P = -2.07+1.94 \log(R/R_\odot)-0.90\log(M/M_\odot).
 \end{array}
\end{equation}
When $P \geq 500$ days, the steady super-wind phase is modeled by
the law
\begin{equation}
\dot{M}(M_\odot {\rm yr}^{-1})=2.06\times
10^{-8}\frac{L/L_\odot}{v_\infty},
\end{equation}
where $v_\infty$ is the terminal velocity of the super-wind in km
s$^{-1}$. We use $v_\infty$=15 km s$^{-1}$ in this paper.\\
(ii)Based on the simulations of shock-driven winds in the
atmospheres of Mira-like stars in \citet{b88}, \citet{b95} gave a
mass-loss rate similar to Reimers' formula:
\begin{equation}
\dot{M}=4.83\times10^{-9}M^{-2.1}L^{2.7}\dot{M}_{\rm Reimers},
\label{eq:bml}
\end{equation}
where $\dot{M}_{\rm Reimers}$ is given by Eq.(\ref{eq:rml}) but
$\eta=0.02$ \citep{sj07}.

On other stellar evolutionary phase, the mass-loss rates are given
by Reimers' formula \citep{r75}:
\begin{equation}
\dot{M}=-4.0\times10^{-13}\eta\frac{LR}{M}{\rm M_\odot yr^{-1}},
\label{eq:rml}
\end{equation}
where $L$, $R$, $M$ are the stellar luminosity, radius and mass in
solar units and we use $\eta$=0.5.

\subsection{Comparison with previously published yields by other models }
In order to test our synthetic model of asymptotic giant branch
stars, we compare the stellar yields in our model with previously
published yields of \cite{vg97}, \cite{m01} and \cite{i04}.
According to the definition of stellar yield in \cite{i04}, we write
the total yield of isotope $j$ by
\begin{equation}
Y_{j}=\int^{\tau(M_{\rm i})}_{0}(X_{j}-X^0_{j})\frac{{\rm d}M}{{\rm
d}t}{\rm d}t ,\label{eq:yield}
\end{equation}
where $\tau(M_{\rm i})$ is the entire lifetime of a star with
initial mass $M_{\rm i}$, $\frac{{\rm d}M}{{\rm d}t}$ is the current
mass-loss rate, $X_{j}$ and $X^0_{j}$ refer to the current and
initial surface abundance of the isotope $j$, respectively.

\begin{figure}
\includegraphics[totalheight=3.in,width=2.8in,angle=-90]{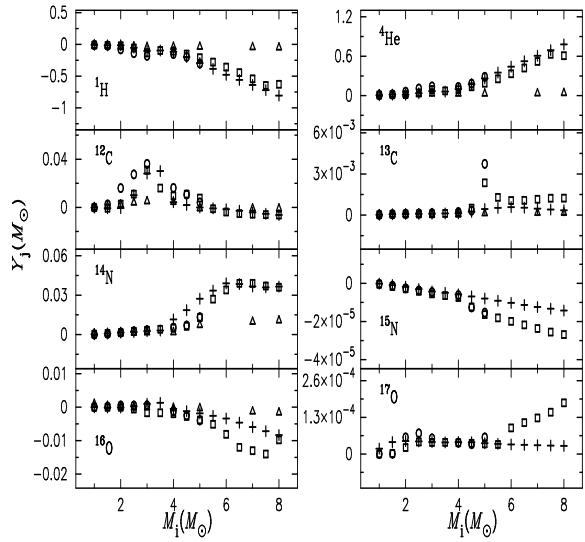}
\caption{Total stellar yields vs. the initial stellar masses.
Triangles, circles and squares represent the models of \cite{vg97},
\cite{m01} and \cite{i04}, respectively. Pluses are our results. }
\label{fig:yield}
\end{figure}

Figure \ref{fig:yield} shows the stellar yields of \cite{vg97} [for
the models of $Z=0.02$, $\eta_{\rm AGB}=4.0$ and $m_{\rm HBB}=0.8$
(See Table 17 of \citealt{vg97})], \cite{m01}(for the models of
$Z=0.019$ and the mixing length parameter $\alpha=1.68$),
\cite{i04}[for the models of $Z=0.02$ (See Table D2 of
\citealt{i04})] and ours. In our models, all physical parameters are
the same as those in case 1 except that we change binary systems
into single star. As Figure \ref{fig:yield} shows, the stellar
yields of our models in general are in agreement with those in
\cite{i04} or lie between those of \cite{vg97} and \cite{i04}
although our models produce more $^{15}$N, and less $^{17}$O at
massive stars.

\label{lastpage}

\end{document}